\documentclass[9pt,a4paper]{article}
\usepackage{jheppub}
\usepackage{lipsum}
\usepackage{braket}
\usepackage{tensor}
\usepackage{dsfont}
\usepackage{tcolorbox}
\usepackage[normalem]{ulem}
\usepackage{caption}
\usepackage{subcaption}
\usepackage{bm}
\usepackage{caption}
\usepackage{multirow}
\usepackage{subcaption}
\usepackage{ragged2e}
\usepackage{cancel}
\usepackage{comment}
\usepackage{xcolor}
\DeclareCaptionJustification{justified}{\justifying}
\usepackage{hyperref}
\usepackage{amssymb, mathrsfs}
\usepackage{bm}
\usepackage{csvsimple}
\usepackage{soul}

\newcommand{\dd}{\mathrm{d}}

\newcommand{\NOm}[2]{:{#2}:_{#1}}
\newcommand{\NO}[1]{:{#1}:}
\newcommand{\gmax}{\sqrt{8\pi}}
\newcommand{\AScomment}[1]{}

\title{Variational Method in Quantum Field Theory}

\author[a]{Arthur Hutsalyuk,}
\author[b]{M\'arton L\'ajer,}
\author[a]{Giuseppe Mussardo}
\author[a] {and Andrea Stampiggi}
\affiliation[a]{
 SISSA \& INFN, Sezione di Trieste, via Bonomea 265, I-34136, Trieste, Italy
}
\affiliation[b]{Division of Condensed Matter Physics and Materials Science, Brookhaven National Laboratory, Upton, NY 11973-5000, USA}

\emailAdd{astampig@sissa.it}

\preprint{}
\abstract{

We develop a variational framework for addressing two-dimensional non-integrable quantum field theories through the exact structure of their integrable counterparts. Concentrating on the $\varphi^4$ Landau–Ginzburg model, we use the analytical Vacuum Expectation Values and Form Factors of local operators in the sinh–Gordon theory as the foundation of a variational ansatz. In this way, we obtain controlled estimates of central physical quantities of the $\varphi^4$ theory--such as the finite-volume ground-state energy and the physical mass as a function of the coupling constant. The strengths of the variational methods are leveraged in combination with the Hamiltonian truncation techniques and the LeClair-Mussardo formula, which also allow to probe the accuracy of the variational approximation varying the system size. Within the weak-coupling regime, a detailed numerical analysis reveals the behaviour of the finite-volume spectrum, the ground-state energy, and the elastic part of the scattering matrix, showing how the rigorous machinery of integrable models can serve as a guiding light into the complex landscape of non-integrable quantum field dynamics.}

\date{\today}
\notoc

\begin{document}
    \maketitle
	\flushbottom
\newpage

\section{Introduction\label{Variational}} 

Despite almost a century of remarkable discoveries, quantum physics remains a domain of intense exploration, largely because exact solutions are seldom attainable within the current mathematical framework. This limitation has elevated the role of approximate methods, among which the variational approach stands as one of the most powerful and versatile tools in modern theoretical physics. Originally a cornerstone technique of quantum mechanics \cite{Sakurai, Shankar}, the variational method has, in recent years, found wide application in quantum many-body systems of reduced dimensionality \cite{Claeys, Claeys-thesis, DePalo} and in quantum field theories \cite{Cirac, Rovira, Guida, VERSCHELDE1992133, Yee:1997fb, Rutkevich_1994}. These low-dimensional quantum field theories are of enduring interest for both fundamental and pragmatic reasons. On one hand, they are testbeds for ideas and mechanisms that also underlie higher-dimensional, more intricate phenomena. On the other, they are fascinating in their own right, often displaying unexpected structures and strikingly nontrivial physics. Thus, the pursuit of variational methods tailored to low-dimensional systems is driven by both extrinsic motivations—testing broader theoretical principles—and intrinsic ones—uncovering the unique beauty and complexity of these models themselves.

In this work, we focus on $(1+1)$-dimensional quantum field theories, a remarkably rich arena where extensive analytical control can often be achieved. Within this class, certain theories admit a complete description in terms of their mass spectrum and two-body scattering matrix. In these integrable models, the dynamics is fully elastic and multi-particle amplitudes factorize into sequences of two-body scatterings — see, for instance, \cite{Zamolodchikov:1989fp, Zamolodchikov:1992zr,Mussardo:2020rxh}. 
A distinguished subset of integrable theories can be obtained as relevant perturbations of conformal field theories \cite{Belavin:1984vu, fateevCONFORMALFIELDTHEORY1990}. For these models, much of the physical content is known exactly and analytically: mass ratios \cite{Fateev:1993av, Zamolodchikov:1995xk}, elastic scattering amplitudes \cite{Zamolodchikov:1989fp, Cardy:1989fw, Christe:1989ah, Sotkov:1989zn, Freund:1989jq,Mussardo:1992uc,Smirnov:1991uw, Smirnov:1990vm,Colomo:1991gw,Zamolodchikov:1991vx,Zamolodchikov:1978xm}, form factors of local operators \cite{Berg:1978sw, Karowski:1978vz, Smirnov_book, Cardy:1990pc, Delfino:1995zk, Delfino:1994ea}, and vacuum expectation values of relevant fields \cite{Lukyanov:1996jj, Fateev:1997yg}. Furthermore, finite-volume quantities can be computed with great precision using numerical methods exploiting integrability \cite{Zamolodchikov:1989cf, Klassen:1990dx,Zamolodchikov:1991vg,Leclair:1999ys}.

By contrast, our understanding of non-integrable theories beyond some perturbative or Bethe-Salpeter approximation \cite{Fonseca:2001dc,Zamolodchikov:2011wd,mvf7-pd7p,Rutkevich:2005ai},  
semiclassical methods \cite{Mussardo:2003ji,Mussardo:2006iv}
or Form Factor Perturbation Theory \cite{Delfino:1996xp,Delfino:1997ya,Delfino:2005bh,PhysRevLett.92.021601}, remains fragmentary, although significant insight can be gained through numerical studies of their spectra, particularly via Hamiltonian truncation techniques \cite{Yurov:1989yu, Yurov:1991my, Lassig:1990xy, Delfino:1996xp, Fitzpatrick:2022dwq, 2015NuPhB.899..547K, 2021JHEP...01..014K, Coser:2014lla, konikReviewTSA, LajerKonik, Horvath:2022zwx, Takacs2012, BERIA2013457, BAJNOK201493,Xu:2022mmw,Xu:2023nke,Brandino:2010sv} and more recently with variational continuous tensor-network methods \cite{tilloy1, tilloy2letter, tilloy2}.

A paradigmatic example is the $\varphi^4$ Landau–Ginzburg theory, the simplest non-integrable scalar field theory. Although it is a textbook prototype for perturbation theory, its physics extends far beyond the perturbative domain. Already in the 1970s, Chang \cite{Chang:1976ek} demonstrated that the $\varphi^4$ model exhibits a second-order phase transition \cite{Simon:1973yz,Glimm:1975tw}: as the coupling varies, the Hamiltonian describing a single-vacuum phase continuously transforms into that of a spontaneously broken one.
Yet, many fundamental aspects of the $\varphi^4$ theory remain elusive. For instance, the exact mass formula is still unknown, despite sustained efforts to refine perturbative predictions \cite{Serone1}. Another key observable is the finite-volume ground-state energy, corresponding to the theory compactified on a cylinder of finite radius.
Intriguingly, both quantities are known exactly in a closely related theory — the sinh–Gordon model. In their symmetric phases, the two theories display striking parallels: each possesses a single neutral excitation above the vacuum, no additional bound states, and an underlying ${\mathbb Z}_2$ invariance. Even more fascinatingly, the Chang singularity of the $\varphi^4$ theory at strong coupling finds a counterpart in the peculiar behavior of the sinh–Gordon model near its self-dual point \cite{2021JHEP...01..014K}.

The central aim of this work is to develop a variational approach that employs the Hamiltonian density of the $\varphi^4$ theory as the key observable, enabling us to estimate physical quantities in the non-integrable $\varphi^4$ model by exploiting exact relations known from the integrable sinh–Gordon theory. We then compare these variational estimates with state-of-the-art numerical and analytical techniques, including the Thermodynamic Bethe Ansatz (TBA) \cite{Zamolodchikov:1989cf}, the Truncated Space Method (TSM) \cite{BajnokMarton2, Coser:2014lla,EliasMiroNLO, EliasMiro:2021aof, Rychkov:2014eea, Rychkov:2015vap}, and the most refined perturbative computations available \cite{Serone1}.

In conventional quantum field theory, Feynman perturbation theory remains one of the most robust and conceptually transparent tools for addressing interacting systems. In the absence of interactions, quantum fields describe free particles, and when interactions are introduced adiabatically, approximating the full interacting fields by their free counterparts is both mathematically consistent and physically justified. This framework allows one to compute, within a unified formalism, corrections to inelastic cross sections and time-ordered correlation functions of local observables. At a deeper level, perturbation theory encodes a correspondence between the Hilbert spaces of the interacting and non-interacting theories. Starting from a Lagrangian $\mathscr{L}$ involving a set of quantum fields, one constructs the corresponding Hamiltonian $\mathscr{H}$, which can typically be decomposed into a free and an interaction part. The key assumption underlying perturbation theory is that the vacuum state of the interacting theory differs from that of the free theory only by analytic corrections in the coupling constants.

However, in other quantum systems—such as those encountered in quantum mechanics or quantum many-body physics—different approximation schemes naturally arise. In these contexts, the interacting system is often modeled variationally, using trial states inspired by the exact spectra of related solvable models. Such variational methods frequently yield quantitative predictions for low-lying excitations and ground-state energies, although they may occasionally fail to capture the qualitative behavior of more complex observables \cite{Sakurai}. In quantum mechanics, the variational principle follows from a fundamental property of Hamiltonians bounded from below: for any family of known Hamiltonians $H(\lambda)$ with ground states $\ket{0_\lambda}$ and eigenvalues $E(\lambda)$, and for an unknown Hamiltonian $H_{\text{int}}$ with ground state energy $E_0$, the following inequality holds
\begin{equation}
\braket{0_\lambda | H_{\text{int}}| 0_\lambda} \geq E_0\,\,\,.
\end{equation} 
The best approximation to the true ground state of $H_{\text{int}}$ is thus obtained by seeking the state $\ket{0_\lambda}$ which minimizes $ \braket{0_\lambda | H_{\text{int}}| 0_\lambda}$.
This procedure, simple in form yet profound in scope, lies at the foundation of variational methods across all of quantum theory.
 
 An analogous principle holds for quantum many-body systems. When such a system is viewed as a statistical ensemble, the variational inequality relates the free energy $F_{\text{int}}$ of the system to be approximated to a family of known free energies $F_0(\lambda)$ corresponding to auxiliary systems whose expectation values $\braket{\ldots}\lambda$ can be computed explicitly. The relation between the two reads \cite{feynman1998statistical}:
\begin{equation}
F_{\text{int}} \geq F_0(\lambda) + \braket{H_{\text{int}} - H_0(\lambda)}_\lambda,
\end{equation}
where $H_{\text{int}}$ and $H_0(\lambda)$ are the Hamiltonians of the interacting and reference systems, respectively.

Extending such variational reasoning to quantum field theory (QFT) presents nontrivial conceptual challenges. In general, the results of quantum mechanics or many-body physics cannot be straightforwardly imported into QFT, since states with different particle numbers are dynamically connected by inelastic processes. This feature also invalidates the thermodynamic analogy: no genuine free energy can be defined. The formal analogue, the generating functional, while useful for computations, is not an observable quantity.
Nevertheless, a physically meaningful variational framework for QFT can be motivated on heuristic grounds. Suppose we wish to estimate the vacuum energy and related observables, such as the mass of the lowest excitation, in a given interacting theory. One may then search for a family of reference theories sharing the same essential features: identical internal symmetries, vacuum degeneracy, and particle content.

For instance, in the $(1+1)$-dimensional $\varphi^4$ theory in the unbroken phase, there exists a single vacuum and a single neutral excitation without bound states. These are precisely the characteristics of the sinh–Gordon model, a one-parameter family of integrable QFTs depending on the coupling $b$, for which the vacuum expectation values of the fundamental scalar operator are exactly known \cite{Fateev:1997yg}. Before performing a variational minimization, it is essential to ensure that the quantities involved are separately finite. Both the vacuum energy and field fluctuations diverge in general. However, for scalar field theories in $(1+1)$ dimensions without derivative interactions, these divergences can be simultaneously renormalized by imposing normal ordering on the free Hamiltonian \cite{Coleman}. By setting the vacuum energy of the free field to zero, the corresponding vacuum energy densities of the $\varphi^4$ and sinh–Gordon models—denoted respectively by $\mathcal{E}_{\varphi^4}$ and $\mathcal{E}_{\text{sh-G}}(b)$—become finite.

Let $\mathscr{H}_{\varphi^4}$ and $\mathscr{H}_{\text{sh-G}}(b)$ denote the Hamiltonian densities of the two theories. Both are regarded as perturbations of the same free theory, sharing the same bare mass and kinetic operator. Within this setup, the variational inequality takes the form
\begin{equation}\label{eq_general_variational}
E_{\varphi^4} \leq E_{\text{sh-G}}(b) + \braket{0_b | \mathscr{H}_{\varphi^4} - \mathscr{H}_{\text{sh-G}}(b) | 0_b},
\end{equation}
where $\braket{0_b|\ldots|0_b}$ denotes the expectation value in the sinh–Gordon vacuum. This expression encapsulates the essence of our approach: a variational estimation of the $\varphi^4$ vacuum energy obtained by evaluating its normal-ordered Hamiltonian on the ground states of the sinh–Gordon model.

The structure of the paper is as follows.
In Section~\ref{s_rev_phi4}, we review the fundamental properties of the $\varphi^4$ theory, while Section~\ref{s_rev_shG} is devoted to the corresponding features of the sinh–Gordon model.
In Section~\ref{s_variational}, we formulate and analyze the variational method in infinite volume.
In Section~\ref{s_TBA}, we provide an estimate of the $\varphi^4$ ground-state energy by combining the Thermodynamic Bethe Ansatz (TBA) with the LeClair–Mussardo series.
Section~\ref{s_TSM} presents numerical results obtained via the Truncated Space Method (TSM) applied to the $\varphi^4$ Hamiltonian, using the sinh–Gordon basis as a variational reference.
For completeness, the TSM algorithm is outlined in Appendix~\ref{a_TSM_explanation}, and our main conclusions are summarized in Section~\ref{s_conclusions}.

\section{Overview of the $\varphi^4$ Model}\label{s_rev_phi4}

The Euclidean $\varphi^4$ theory with bare mass $m>0$ and coupling $g \geq 0$ is defined by the Hamiltonian density
\begin{equation}
\label{eq_phi4_hamiltonian}
\mathscr{H}_{\varphi^4}(x) =
\frac{1}{2}\,\pi_0^2(x)
+ \frac{1}{2}\,(\partial_1 \varphi)^2(x)
+ \frac{m^2}{2}\,\varphi^2(x)
+ m^2 g\,\varphi^4(x).
\end{equation}
Here $\pi_0$ denotes the canonical momentum conjugate to the field $\varphi$, that is, $\pi_0 = \partial_0 \varphi$. In the Euclidean formulation, the Lagrangian density coincides with the Hamiltonian, whereas in Minkowski space-time the Lagrangian differs by relative minus signs in the last three terms. 

The first three contributions in \eqref{eq_phi4_hamiltonian} describe the free evolution of a massive scalar field, while the last term encodes the quartic self-interaction. Since the impact of this interaction on the particle spectrum is a priori unknown, the natural starting point is to adopt as a variational basis the free-particle states of the corresponding non-interacting theory. The most consistent choice is to take the bare mass $m$ appearing in the Hamiltonian as the reference mass. The single-particle energy is then given by
\begin{equation}
\omega_{p;m} = \sqrt{p^2 + m^2}\,\,\,,
\end{equation}
$p$ being the spatial momentum. For $(1+1)$-dimensional theories, it is convenient to employ the rapidity mapping:
\begin{equation}\label{eq_rapidity}
    p = m \sinh \theta , \quad \omega_{p;m} = m \cosh\theta.
\end{equation}
which linearizes Lorentz boosts and greatly simplifies the kinematic analysis.

In what follows, we shall consider the $\varphi^4$ theory in Minkowski space-time. In the absence of the self-interaction term, the field is free and can be expressed as a linear superposition of creation and annihilation operators acting on the vacuum, each associated with a particle of mass $m$.
\begin{equation}\label{eq_field_decomposition_modes}
   \begin{aligned}
    \varphi\left(x,t\right) &= \int_{-\infty}^\infty \frac{\dd p}{\sqrt{2\pi}} \frac{1}{\sqrt{2\omega_{p;m}}}\left(a_{p;m} e^{-i\omega_{p;m} t} e^{i p x} + a_{p;m}^\dagger e^{i\omega_{p;m} t}e^{-i p x}\right),\\
    \pi_0 (x,t) &= i \int_{-\infty}^\infty \frac{\dd p}{\sqrt{2\pi}} \sqrt{\frac{\omega_{p;m}}{2}}\left(a_{p;m} e^{-i\omega_{p;m} t} e^{i p x} - a_{p;m}^\dagger e^{i\omega_{p;m} t}e^{-i p x}\right).
   \end{aligned}
\end{equation}
For an interacting quantum field theory, the free-field representation of the field operators remains exact only in the asymptotic regime, where particle states are infinitely separated in space-time. At finite separations, this representation ceases to hold, as interactions become non-negligible.

When expressing the Hamiltonian in terms of creation and annihilation operators, two distinct divergences naturally arise: one from the continuum nature of the spectrum and the other originating from contributions at ultraviolet momenta. The first divergence can be regulated by confining the system within a finite spatial volume, while the second may be controlled by discretizing space on a finite lattice. An elegant alternative is to impose normal ordering of the field operators, which systematically removes such divergences when acting on physical states. In a string of operators, annihilation operators are placed to the right of the creation operators, ensuring that all vacuum contractions vanish.
Normal ordering is always defined with respect to a particle basis of a given mass. Operationally, it eliminates the divergences of the free Hamiltonian—such as those appearing in the vacuum expectation value of the energy for particles of mass $m$, denoted $\ket{0_m}$. To illustrate this, consider the divergence associated with the mass term, obtained via point splitting:
\begin{equation}\label{eq_div_two_point}
\begin{aligned}
\braket{0_m|\varphi(x,0)\varphi(0,0)|0_m}
&= \frac{1}{4\pi} \int_{-\infty}^{\infty}  \dd p  \frac{e^{-ipx}}{\omega_{p;m}}
= \frac{1}{2\pi} K_0 (m x) \\
&\underset{x \to 0}{\simeq} \frac{1}{2\pi}
\left[-\ln (m x) - \ln\left(\frac{e^{\gamma_E}}{2}\right) + O((m x)^2)\right],
\end{aligned}
\end{equation}
where $K_0(z)$ is the modified Bessel function and $\gamma_E$ is the Euler–Mascheroni constant. Usually, one chooses $m$ as the normal-ordering mass, but in general it can be any mass $\mu$. The normal ordering with respect to a different mass is equivalent to shifting the quadratic (mass) term by a finite counterterm. The procedure effectively redefines the vacuum to $\ket{0_\mu}$ and alters the oscillator content of the field.

As shown by Coleman \cite{Coleman}, in the $\varphi^4$ model—and indeed in any scalar field theory without derivative interactions—normal ordering simultaneously removes the ultraviolet divergences from other observables and from the $S$-matrix in the interaction picture. In perturbation theory, such divergences originate from self-contractions of the field inside internal loops.
For instance, the first-order correction to the vacuum energy density arises from the expectation value
$\braket{0_m | g \varphi^4(x) | 0_m}$,
where four particles are created at the same space-time point. Two pairs can form with opposite momenta, yielding the divergent double-tadpole diagram:
\begin{equation}
\centering
\begin{gathered}
\includegraphics[width=0.25\textwidth]{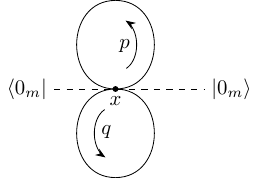}
\end{gathered}
\quad =
6 m^2 g
\int_{-\infty}^\infty \frac{\dd p}{2\pi} \frac{1}{2\omega_{p;m}}
\int_{-\infty}^\infty  \frac{\dd q}{2\pi} \frac{1}{2\omega_{q;m}}.
\end{equation}
These integrals can be regularized either by introducing an ultraviolet cutoff $\Lambda$ or, equivalently, by applying an infinitesimal point splitting around $x$. Each integral displays a leading logarithmic divergence of the same form as in Eq.~\eqref{eq_div_two_point}. Consequently, one of them can be associated with the renormalization of $\varphi^2$, while the remaining finite part depends only on the regularization constant.

The mass counterterm can then be fixed to cancel the divergent vacuum contribution exactly:
\begin{equation}
\mu^2 - m^2 =
6\frac{m^2 g }{\pi}
\log \left[
\frac{\Lambda}{m}
\left(1 + \sqrt{1 + \left(\frac{m}{\Lambda}\right)^2}\right)
\right].
\end{equation}
This procedure extends naturally to higher-order corrections of the vacuum energy, yielding a theory free from divergent self-interaction energies. Crucially, the same divergences appear for multi-particle states, but they are of identical origin—tadpole diagrams—and are absorbed by the same mass counterterm.

Thus, the argument generalizes to all local observables constructed from the fundamental field $\varphi$ and to the $S$-matrix elements. The same regularization and renormalization scheme applies to any scalar theory without derivative interactions, and in particular, to the sinh–Gordon model as well.
The generality of the normal-ordering regularization allows one to establish universal relations connecting the same scalar theory—provided it contains no derivative interactions—under different normal-ordering prescriptions.
Let $\mathscr{H}_m$ denote the Hamiltonian density with bare mass parameter $m$, and let
$\NOm{\mu}{\mathcal{O}}$ represent the normal ordering of a local operator $\mathcal{O}$, constructed from the fundamental field $\varphi$, with respect to a reference mass $\mu$.
As shown in \cite{Coleman}, the following exact relations hold:
\begin{equation}\label{eq_normal_ordering_scalar}
    \begin{aligned}
         \NOm{\mu_1}{\mathscr{H}_m} &= \NOm{\mu_2}{\mathscr{H}_m} + \frac{1}{8\pi} (\mu_1^2-\mu_2^2) - \frac{1}{4\pi}m^2 \ln\frac{\mu_1}{\mu_2}, \\
         \NOm{\mu_1}{\cos\beta\varphi} &=\left(\frac{\mu_2^2}{\mu_1^2}\right)^{\frac{\beta^2}{8\pi}}\NOm{\mu_2}{\cos\beta\varphi}, \\
         \NOm{\mu_1}{\cosh b \varphi} &=\left(\frac{\mu_2^2}{\mu_1^2}\right)^{-\frac{b^2}{8\pi}}\NOm{\mu_2}{\cosh b \varphi}.
    \end{aligned}
\end{equation}
These relations are basis-independent and stem purely from the field-theoretic structure of the theory.

It is particularly noteworthy that, through these normal-ordering identities, one can explicitly demonstrate the equivalence—at the level of normal-ordered Hamiltonians—between the $\varphi^4$ theory at strong coupling in the symmetric phase and the same theory at weak coupling in the broken phase. This remarkable correspondence, known as Chang duality \cite{Chang:1976ek}, highlights a deep self-consistency within the scalar field dynamics. 

In the present work, we focus on the $\varphi^4$ theory in the symmetric, weak-coupling regime, where the spectrum consists of a single bosonic excitation without bound states.
Regarding perturbative corrections to the vacuum energy density $\mathcal{E}_{\varphi^4}$ and the squared physical mass $M_{\varphi^4}^2$, we recall the results obtained in \cite{Serone1}, valid up to order $g^8$:
\begin{equation}\label{eq_vacuum_mass_phi4_perturbative}
    \begin{aligned}
    \frac{\mathcal{E}_{\varphi^4}}{m^2}= & -\frac{21 \zeta(3)}{16 \pi^3} g^2+\frac{27 \zeta(3)}{8 \pi^4} g^3-0.116125964(91) g^4+0.3949534(18) g^5 + \\
    &\quad -1.629794(22) g^6+7.85404(21) g^7-43.1920(21) g^8+\mathcal{O}\left(g^9\right),\\
    \frac{M^2_{\varphi^4}}{m^2}= & 1-\frac{3}{2} g^2+\left(\frac{9}{\pi}+\frac{63 \zeta(3)}{2 \pi^3}\right) g^3-14.655869(22) g^4+65.97308(43) g^5 + \\
    &\quad -347.8881(28) g^6+2077.703(36) g^7-13771.04(54) g^8+\mathcal{O}\left(g^9\right) .
\end{aligned}
\end{equation}
These expansions are asymptotic, but they can be rendered predictive by applying Borel resummation techniques, as discussed in \cite{Serone1}. The resulting Borel-resummed series will serve as the primary analytical benchmark for comparison with our variational estimates, which, as we shall see, are reliable for $g \leq 1/3$.

\section{Overview of the sinh-Gordon Model}\label{s_rev_shG}

The sinh-Gordon model in $(1+1)$-dimensions is an integrable quantum field theory -- see \cite{Mussardo:2020rxh} for a textbook introduction -- of Hamiltonian density
\begin{equation}\label{eq_shG_hamiltonian}
    \mathscr{H}_{\text{sh-G}} (x) = \frac{1}{2} \pi_0^2(x) + \frac{1}{2}(\partial_1 \varphi)^2 (x) + \frac{m^2}{ b^2}\left(\cosh(b\varphi)-1\right).
\end{equation}
A direct comparison with the $\varphi^4$ Hamiltonian density in Eq.~\eqref{eq_phi4_hamiltonian} immediately reveals that the two energy densities, when expanded around the classical stationary configuration $\varphi = 0$, coincide up to order $\varphi^4$, provided one identifies the parameters through
 $b^2 = 4! g$. 
 This correspondence, however, holds strictly at the classical level and within the perturbative regime of the quantum theory; beyond that, quantum corrections spoil the equivalence, as will later emerge from the variational analysis.
 
Thanks to the remarkable simplicity of the sinh–Gordon spectrum, much of its structure can be determined exactly and analytically. Owing to its integrability, all scattering processes are purely elastic and the multi-particle amplitudes factorize into products of two-body interactions. The elastic $S$-matrix, expressed in terms of the rapidity difference $\theta = \theta_2 - \theta_1$, is given by \cite{koubek1993operator}
\begin{equation}\label{eq_shG_smatrix}
    \braket{\theta_1, \theta_2| S | \theta_1, \theta_2} = S_b(\theta) = \frac{\tanh \frac{1}{2}(\theta - i \pi\alpha)}{\tanh \frac{1}{2}(\theta + i \pi \alpha)} = \frac{\sinh\theta - \sinh(i\pi \alpha)}{\sinh\theta + \sinh(i\pi \alpha)}, \quad \alpha = \frac{b^2}{b^2 + 8 \pi}.
\end{equation}
The $S$-matrix presents a duality in the mapping $b \rightleftharpoons 8\pi/b$, corresponding to $\alpha \rightleftharpoons (1-\alpha)$, identifying $b^\star = \sqrt{8\pi}$ as the ``self-dual'' point. The nature of the sinh-Gordon theory at the self-dual point and beyond is a problem still under debate, see \cite{2021JHEP...01..014K, bernard2022sinh} for detailed and in-depth analyses. No poles in the physical strip $0 \leq \operatorname{Im} \theta <\pi$ are present, as such there is only one particle excitation without bound states. The mass $M_{\text{sh-G}}(b)$ of such particle reads \cite{Zamolodchikov:1995xk,Lukyanov:1996jj,Fateev:1997yg}\footnote{Let us note that the result in the literature actually differ in the Hamiltonian convention. We will denote their quantities by a $\tilde a$, $\tilde b$. First, the kinetic term is rescaled by a factor $1/8\pi$, or equivalently that $\tilde{\varphi} = \sqrt{8\pi}\varphi$. As such, the coupling $\tilde{b} = b/\sqrt{8\pi}$, i.e. the self-dual point is now found at $\tilde{b}^\star = 1$. The bare coupling multiplying the $\cosh$ is denoted by $2\tilde{\mu} = m^2/(8\pi \tilde{b}^2)$.}:
\begin{equation}\label{eq_shG_mass_squared}
    M^2_{\text{sh-G}}(b) = \frac{16 \pi}{\left[\Gamma\left(\frac{1}{2+2\tilde{b}^2}\right)\Gamma\left(1+\frac{\tilde{b}^2}{2+2\tilde{b}^2}\right)\right]^2}\left[-\tilde{\mu}_{UV} \pi \gamma(1+\tilde{b}^2)\right]^{\frac{1}{1+\tilde{b}^2}},
\end{equation}
where $\tilde{b} = b/\sqrt{8\pi}$ and  $\gamma(z) = \Gamma(z)/\Gamma(1-z)$. Here $\tilde{\mu}_{UV} = \left[m^2/(16\pi \tilde{b}^2)\right] \left(m e^{\gamma_E}/2\right)^{2\tilde{b}^2}$ is the ultraviolet-finite part of the sinh-Gordon coupling, which as well as the sine-Gordon theory, undergoes multiplicative renormalization \cite{Coleman}. 

The sinh-Gordon model can be seen as a deformation of the Gaussian fixed point, where the exponential field has an operator product expansion on the massless vacuum \cite{2021JHEP...01..014K}:
\begin{equation}
    \braket{\cosh(b\varphi(x)) \cosh(b\varphi(0))} \simeq \frac{\braket{\cosh(2b\varphi(0))}}{\lvert x \rvert^{4 b^2/8\pi}} + \lvert x \rvert^{4 b^2/8\pi} + \ldots.
\end{equation}
It is customary to adopt the ``conformal normalization'' for the expectation values of the field, in which we require that $\braket{\varphi^n}_{\text{C}} = 0$ on the massless vacuum. The relation between the massless and massive bases has been worked out, see e.g. Appendix A of \cite{2021JHEP...01..014K}.

Since our goal is to establish a direct relation between the couplings $g$ and $b$ of the $\varphi^4$ and sinh–Gordon theories, respectively, we first need to match the two models within a common quantization prescription. To this end, we fix the sinh--Gordon model in the same massive normal-ordering scheme used for the $\varphi^4$ Hamiltonian: all local operators are normal ordered with respect to the free massive boson of mass $m$, and we denote this prescription by $:\,\cdots\,:_m$.
It is convenient to introduce the point-splitting scale 
\[
m_{\rm ps}\equiv m\,\frac{e^{\gamma_E}}{2}\,,
\]
associated with the constant prefactor appearing in the short-distance expansion of the two-point function eq. \eqref{eq_div_two_point}.

When quoting vacuum expectation values of exponential fields, we will package the canonical mass dimension using $m_{\rm ps}$ and define the dimensionless quantity
\begin{equation}\label{eq_vev_adimensional_shG}
    \big\langle :e^{a\varphi}:\big\rangle_b
    \equiv \left(m_{\rm ps}^2\right)^{a^2/8\pi}\,
    \big\langle :e^{a \varphi}:_{m}\big\rangle_b \, ,
\end{equation}
where $\braket{\NOm{m}{e^{a \varphi}}}$ was derived in \cite{Fateev:1997yg} and reads
\begin{equation}\label{eq_vev_FLZZ_generalformula}
    \braket{\NOm{m}{e^{a \varphi}}}_b = \left(M^2_{\text{sh-G}}\right)^{-\tilde{a}^2} \left(\frac{\Gamma\left(\frac{1}{2+2\tilde{b}^2}\right)\Gamma\left(1+\frac{\tilde{b}^2}{2+2\tilde{b}^2}\right)}{4\sqrt{\pi}}\right)^{-2 \tilde{a}^2} F_b(a),
\end{equation}
where the dependence on $m$ is through {\color{blue}$\tilde{\mu}_{UV}$}\st{$\tilde{\mu}$}, $\tilde{a} = a/\sqrt{8\pi}$ and
\begin{equation}\label{eq_vev_FLZZ_exponential}
    F_b(a) = e^{f_b (a)}, \quad f_b (a) = \int_0^\infty \frac{\dd t}{t} \left(-\frac{\sinh^2\left(2\tilde{a}\tilde{b}t\right)}{2\sinh\left(\tilde{b}^2 t\right) \sinh t \cosh\left[(1+\tilde{b}^2) t\right]} + 2 \tilde{a}^2 e^{-2t} \right). 
\end{equation}

The expression above converges for $\tilde{a} < \frac{1}{2} (\tilde{b}+1/\tilde{b})$, 
so that a cumulant expansion may be employed to extract the moments $\braket{\NO{\varphi^n}}_b$ by taking the vanishing-$a$ limit of Eq.~\eqref{eq_vev_adimensional_shG}. The general formula for the cumulants in the dimensionless scheme reads \cite{Lukyanov:1997bp}:
\begin{equation}\label{eq_cumulant_general_formula}
\begin{aligned}
\sigma_{2n}&=(-1)^n 4^{2 n-1} \tilde{b}^{2 n}  \int_0^{\infty}\frac{\dd t}{t} \frac{t^{2 n}}{\sinh (\tilde{b}^2 t) \sinh (t) \cosh ((1+\tilde{b}^2)t)}+\\
&\quad - \frac{1}{2\pi} \delta_{1,\; n} \log \left(\frac{M_{\text{sh-G}}}{m_{\text{ps}}}\frac{\Gamma\left(\frac{1}{2+2\tilde{b}^2}\right)\Gamma\left(1+\frac{\tilde{b}^2}{2+2\tilde{b}^2}\right)}{4\sqrt{\pi}}\right),
\end{aligned}
\end{equation}
and
\begin{equation}
\exp(a\varphi)=1+\sum_{n=1}^{\infty}\frac{a^{2n}}{(2n)!}\braket{\NO{\varphi^{2n}}}_b=\exp\left(\sum_{n=1}^{\infty}\frac{a^{2n}\sigma_{2n}}{(2n)!}\right).
\end{equation}
Then,
\begin{equation}
\braket{\NO{\varphi^{2}}}_b=\sigma_2,\qquad \braket{\NO{\varphi^{4}}}_b=\sigma_{4}-3\braket{\NO{\varphi^{2}}}_b^2.
\end{equation}

\begin{figure}
    \centering
    \includegraphics[width=0.6\linewidth]{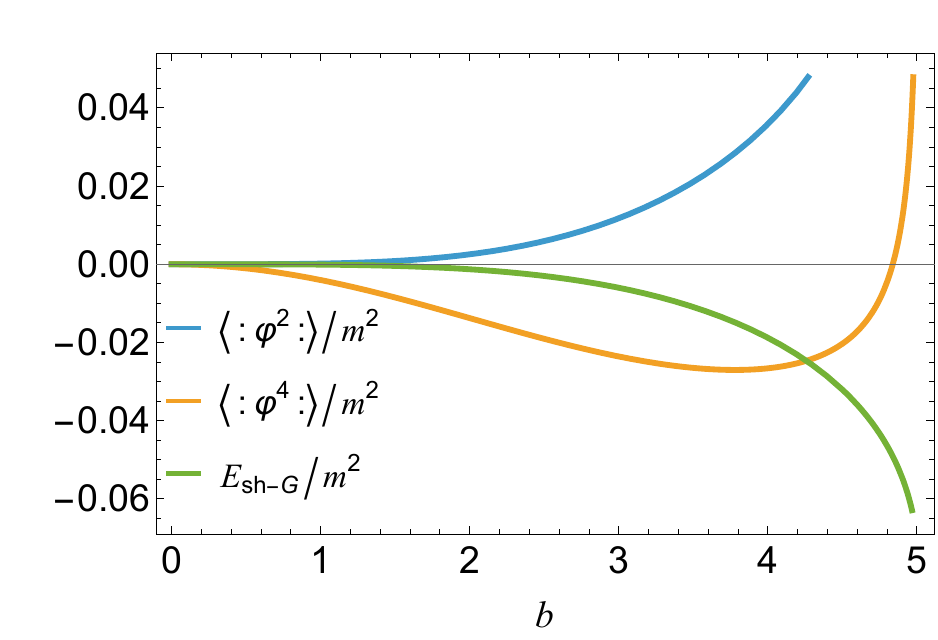}
    \caption{Plot of vacuum expectation values of the field powers, Eq.~\eqref{eq_vev_powers_shG} and of the vacuum energy density Eq.~\eqref{eq_vacuum_energy_density_shG}. In the free limit $b=0$, all plotted quantities vanish in the massive normal ordering scheme. Here $\braket{:\varphi^2:} $ is monotonically increasing, while $\braket{:\varphi^4:}$ is zero at $b\approx 4.83764$. Because of the kinetic contribution, the vacuum energy density is overall monotonically decreasing.}
    \label{f_vevs_vacuum_energy}
\end{figure}

Odd expectation values vanish, since the model has a ${\mathbb Z}_2$ symmetry, and therefore only even expectation values remain. The first two non-vanishing expectation values are:
\begin{equation}\label{eq_vev_powers_shG}
    \begin{aligned}
    \braket{\NO{\varphi^2}}_b 
    &= -\frac{1}{2\pi} \log \left(\frac{M_{\text{sh-G}}}{m_{\text{ps}}}\frac{\Gamma\left(\frac{1}{2+2\tilde{b}^2}\right)\Gamma\left(1+\frac{\tilde{b}^2}{2+2\tilde{b}^2}\right)}{4\sqrt{\pi}}\right)+\\
    &\quad - \frac{1}{2\pi}\int_0^\infty \frac{\dd t}{t}\left[\frac{\tilde{b}^2 t^2}{\sinh(\tilde{b}^2 t)\sinh(t)\cosh((1+\tilde{b}^2)t)}- e^{-2t}\right] \\
    \braket{\NO{\varphi^4}}_b,
    &= -\frac{1}{\pi^2} \int_0^\infty \frac{\dd t}{t}\frac{\tilde{b}^4 t^4}{\sinh(\tilde{b}^2 t)\sinh(t)\cosh((1+\tilde{b}^2)t)} + 3 \braket{\NO{\varphi^2}}_b^2.
\end{aligned}
\end{equation}
The normal ordering ensures that all field powers, as functions of $b$, vanish in the free-field limit $b\to 0$, as can be seen explicitly from the plots in Fig.~\ref{f_vevs_vacuum_energy}. Notice also that $\braket{\NO{\varphi^2}}_b$ is positive and monotonically increasing.

The vacuum expectation value of the potential can be retrieved from Eq.~\eqref{eq_vev_adimensional_shG} when $a$ equals $b$:
\begin{equation}\label{eq_vev_cosh_adimensional_shG}
    \braket{\NO{\cosh b\varphi}}_b = \pi\frac{\tilde{b}^2}{1+\tilde{b}^2} \frac{(M_{\text{sh-G}}/m)^2}{\sin\left(\pi \frac{\tilde{b^2}}{1+\tilde{b}^2}\right)}\,\,\,.
\end{equation}
This expression is related to the expectation value of the trace of the stress-energy tensor coming from conformal perturbation theory \cite{Mussardo:2020rxh} and is also connected to the one in the massive scheme by an overall rescaling of $8\pi$:
\begin{equation}\label{eq_vev_trace_adimensional_shG}
    \braket{\NO{\Theta}}_b =(8\pi) 4\pi (1+\tilde{b}^2) (2\tilde{\mu} )\braket{\NO{\cosh b\varphi}}_b = \frac{\pi}{2} \frac{M_{\text{sh-G}}^2}{\sin\left(\pi \frac{\tilde{b^2}}{1+\tilde{b}^2}\right)},
\end{equation}
which has the correct dimension of $[\text{mass}]^2$.

From the trace of the stress-energy tensor, it is possible to compute the vacuum energy density $\mathcal{E}_{\text{Sh-G}}$, through the thermodynamic Bethe Ansatz (TBA) \cite{Zamolodchikov:1989cf}. If the plane is mapped to a cylinder of finite radius $R$, the ground state energy $E_0 (R)$ will have a leading contribution $R \mathcal{E}_{\text{sh-G}}$ for large enough $R$. Since on a finite radius
\begin{equation}
    \braket{\Theta}_R = 2\pi \frac{1}{R} \frac{\dd }{\dd R} \left(R E_0 (R)\right),
\end{equation}
it follows -- in the limit $R\to\infty$ -- that the vacuum energy density is:
\begin{equation}
\label{eq_vacuum_energy_density_shG_divergent}
    \mathcal{E}_{\text{sh-G}} = \frac{1}{4\pi} \braket{\NO{\Theta}}_b = \frac{1}{8} \frac{M_{\text{sh-G}}^2}{\sin\left(\pi \frac{\tilde{b^2}}{1+\tilde{b}^2}\right)}-\frac{m^2}{8\pi}. 
\end{equation}
The expectation value $\braket{\NO{\mathscr{H}_{\text{sh-G}}}}_b$ is the finite part of $\mathcal{E}_{\text{sh-G}}$, after removing the divergence at $b\to 0$, which exactly equals the term $m^2/b^2$ subtracted in the classical potential term:
\begin{equation}
 \label{eq_vacuum_energy_density_shG}\braket{\NO{\mathscr{H}_{\text{sh-G}}}}_b = \mathcal{E}_{\text{sh-G}} - \frac{m^2}{b^2} = \frac{1}{8} \frac{M_{\text{sh-G}}^2}{\sin\left(\pi \frac{\tilde{b^2}}{1+\tilde{b}^2}\right)}- \frac{m^2}{8\pi} - \frac{m^2}{b^2}.
\end{equation}
From the latter equation, we can retrieve the expectation value for the kinetic term in Eq.~\eqref{eq_shG_hamiltonian}:
\begin{equation}\label{eq_vacuum_kinetic_shG}
    \mathcal{K}_b= \braket{\NO{\frac{1}{2} \pi_0^2(x) + \frac{1}{2}(\partial_1 \varphi)^2 (x)}}_b  =\braket{\NO{\mathscr{H}_{\text{sh-G}}}}_b - \frac{m^2}{b^2}  \braket{\NO{\cosh b\varphi}}_b = \frac{b^2}{8\pi+b^2} \braket{\NO{\mathscr{H}_{\text{sh-G}}}}_b.
\end{equation}

\section{Variational Method at Infinite Volume}\label{s_variational}

Having at our disposal the exact expressions for the vacuum expectation values of the exponential field, we are now in a position to implement the variational principle. As a variational ansatz, we select the sinh–Gordon theory with an arbitrary coupling $b$, and denote by $\ket{0_b}$ the corresponding vacuum state. Our goal is to minimize the vacuum expectation value of the $\varphi^4$ Hamiltonian density, Eq.~\eqref{eq_phi4_hamiltonian}, evaluated on the sinh–Gordon vacuum.

For both the sinh–Gordon and $\varphi^4$ theories, we adopt the same bare mass, and seek to determine the optimal mapping $b(g)$ that minimizes
\begin{equation}
\label{eq_vacuum_energy_phi4}
\braket{0_b | \mathscr{H}_{\varphi^4} | 0_b}
= \mathcal{K}_b
+ \frac{m^2}{2}\,\braket{\NO{\varphi^2}}_b
+ m^2 g\,\braket{\NO{\varphi^4}}_b,
\end{equation}
where $\mathcal{K}_b$ is defined in Eq.~\eqref{eq_vacuum_kinetic_shG}, and the vacuum expectation values of the field powers are given by Eq.~\eqref{eq_vev_powers_shG}.
Since the vacuum energy density of the sinh–Gordon model is known analytically, see Eq.~\eqref{eq_vacuum_energy_density_shG}, the minimization of Eq.~\eqref{eq_vacuum_energy_phi4} is equivalent to the general variational condition in Eq.~\eqref{eq_general_variational}.

\begin{figure}
    \centering
    \includegraphics[width=0.6\linewidth]{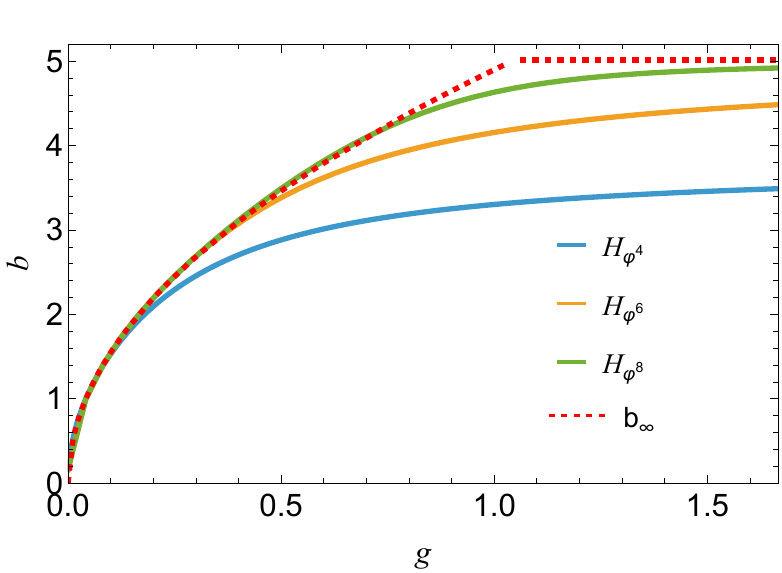}
     \caption{Plot of the function $b(g)$ 
     for various Hamiltonians of type Eq.~\eqref{eq_truncation_shG}, i.e $\mathcal{H}_{4}$ (blue), $\mathcal{H}_{6}$ (orange) and $\mathcal{H}_{8}$ (green). As more powers are considered, the $b(g)$ curves approach Eq.~\eqref{eq_g_star} (red dashed). For $\mathcal{H}_{4}$, whose solution is given formally by Eq.~\eqref{eq_formal_g_lambda}, the agreement with the classical solution is satisfactory for $g\lesssim 1/3$.}
    \label{f_b_lambda}
\end{figure}

In terms of the definitions of the kinetic term and the VEVs, the minimal points of Eq.~\eqref{eq_vacuum_energy_phi4} satisfy
\begin{equation}
\label{eq_formal_g_lambda}
    g(b) = - \frac{1}{\partial_b \braket{\NO{\varphi^4}} }\left(\partial_b \mathcal{K}+ \frac{1}{2}\partial_b \braket{\NO{\varphi^2}}\right)\,\,\,.
\end{equation}
The resulting dependence of $b(g)$ is shown in Fig.~\ref{f_b_lambda}. From the plot, it is clear that the function scales as $\sqrt{24 g}$ in the weak-coupling regime and increases monotonically toward the limiting value $b^\star = \gmax$. This behaviour precisely reproduces the result obtained from a classical expansion of the potential around the stationary configuration $\varphi = 0$. However, since the sinh–Gordon model is an interacting theory, the function $b(g)$ deviates from its classical trend once the coupling exceeds roughly $g \simeq 1/3$.

As $b$ approaches its maximal value $\gmax$, the vacuum expectation values (VEVs) of the exponential field vanish, while those of the fundamental field $\varphi$ diverge. As a consequence, if one considers the sinh–Gordon Hamiltonian density truncated at order $\varphi^{2n}$, namely
\begin{equation}\label{eq_truncation_shG}
    \mathscr{H}_{n} = \frac{1}{2}(\partial_\nu \varphi)(\partial^{\nu}\varphi) + \frac{m^2}{2} \varphi^2 + m^2\sum_{\ell=2}^{n} \frac{(24 g)^{\ell-1}}{2\ell!} \varphi^{2\ell},
\end{equation}
one expects that the optimal mapping $b(g)$, as $n$ increases, should asymptotically approach
\begin{equation}\label{eq_g_star}
    b_{\infty} (g) = \begin{cases}
        \sqrt{24 g}, &\text{if }0\leq g \leq \pi/3,\\
        \gmax, &\text{if } g>\pi/3.
    \end{cases}
\end{equation}
This behavior is clearly confirmed in Fig.~\ref{f_b_lambda}, which displays the results obtained for the truncated Hamiltonians $\mathscr{H}_{4}$, $\mathscr{H}_{6}$, and $\mathscr{H}_{8}$. The trend is evident: as higher-order terms are included, the variational relation $b(g)$ progressively converges toward the asymptotic form in Eq.~\eqref{eq_g_star}. From Eq.~\eqref{eq_g_star} we also deduce the validity domain of this variational method: $g\lesssim 1$, which is well-below the Chang duality point of $\varphi^4$, $g_* = 2.807(34)$ \cite{Serone1}. 

\subsection{Ground state energy\label{GE-phi-4}}

\begin{figure*}
\begin{subfigure}{0.48\linewidth}
\includegraphics[width=\linewidth]{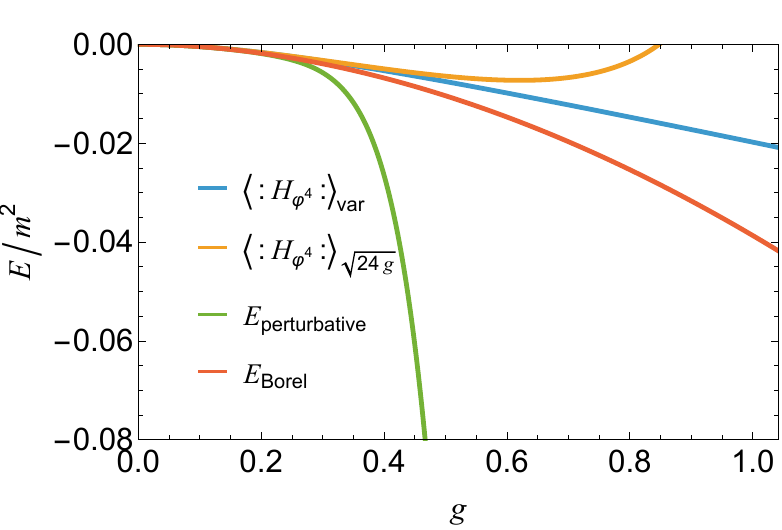}
\subcaption{}
\end{subfigure}
\hfill
\begin{subfigure}{0.48\linewidth}
\includegraphics[width=\linewidth]{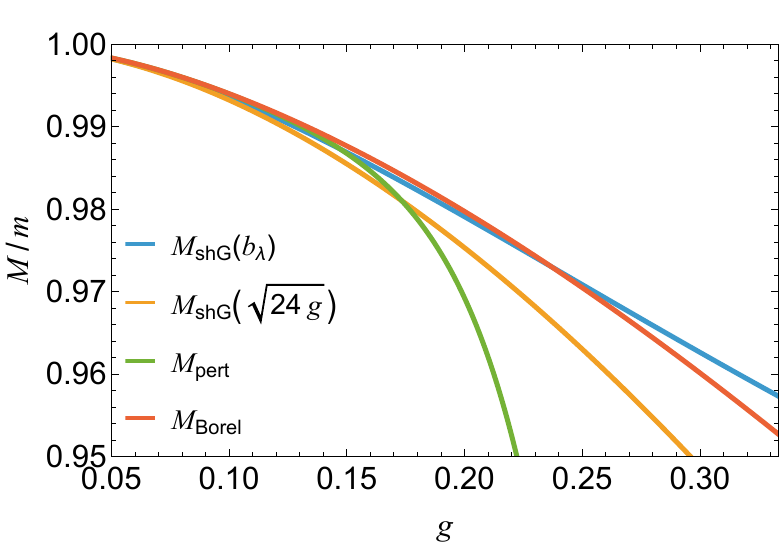}
\subcaption{}
\end{subfigure}%
\caption{a) Plot of the vacuum energy density obtained from Eq.~\eqref{eq_vacuum_energy_phi4} through minimization (blue) and in the leading order approximation $b=\sqrt{24 g}$ (orange). It is also plotted the vacuum energy from  perturbation theory up to $O(g^8)$ Eq.~\eqref{eq_vacuum_mass_phi4_perturbative} (green) and its Borel resummation (red), obtained through the technique of \cite{Serone1}. The energy obtained through the variational principle agrees with the state-of-the-art Borel resummation within $2\cdot 10^{-3}$ for $g \lesssim 1/3$. b) Plot of the sinh-Gordon mass Eq.~\eqref{eq_shG_mass_squared} through the coupling relation Eq.~\eqref{eq_formal_g_lambda}(blue) versus the leading order approximation (orange), in the range $g \leq 1/3$. Here too is shown the estimate from perturbation theory up to $O(g^8)$ (green) and its Borel resummation (red). In the range, the mass estimated obtained through the variational principle agrees with the Borel resummation within $1\cdot 10^{-2}$.}
\label{f_energy_mass}
\end{figure*}

The most significant validation of Eq.~\eqref{eq_formal_g_lambda} comes from the comparison of the predicted vacuum energy density of the $\varphi^4$ theory with state-of-the-art approaches. In particular, we have considered Borel resummation techniques \cite{Serone1} (and, where available, TSM results) applied to the perturbative expansions given in Eq.~\eqref{eq_vacuum_mass_phi4_perturbative}. As shown in Fig.~\ref{f_energy_mass}, the variational prediction provides a markedly improved description of the $\varphi^4$ vacuum energy compared with the classical relation $b(g) = \sqrt{24 g}$, whose behavior is not monotonically decreasing, and with the perturbation theory, Eq.~\eqref{eq_vacuum_mass_phi4_perturbative}, which fails at $g\approx 1/3$.

The minimized sinh-Gordon energy is qualitatively close to the $\varphi^4$ for $g \leq 1/3$, where the maximum deviation is within $2\cdot 10^{-3}$. As argued before, the variational principle using the sinh-Gordon is to be trusted for $g\lesssim \pi/3$, where the classical relation $b(g) = \sqrt{24 g}$ reaches $\gmax$ and the sinh-Gordon VEVs, as well as the ground state energy, deviate from the Borel resummation prediction -- see Fig.~\ref{f_vevs_vacuum_energy}. At larger couplings $1/3\leq g\leq \pi/3$, however, the variational vacuum energy and the one obtained through the Borel resummation deviate considerably, as higher-order field powers VEVs in the sinh-Gordon Hamiltonian become non-negligible compared to $\varphi^4$.

\subsection{Physical Mass \label{Mass-phi-4}}

A particularly nontrivial test of the variational approach is the comparison between observables of the $\varphi^4$ theory, $\mathcal{O}_{\varphi^4}(g)$, with their analogues in the sinh-Gordon theory, $\mathcal{O}_{\text{sh-G}}(b(g))$, computed using the variational relation $b(g)$. The physical mass, the first excitation on top of the vacuum, is a natural candidate. Thus, we compare the Borel-resummed mass $M_{\text{Borel}}(g)$ with $M_{\text{sh-G}}(b(g))$, where $b(g)$ comes from Eq.~\eqref{eq_formal_g_lambda}. For $g\leq 1/3$, they agree within $1\cdot 10^{-2}$. We also compare with the perturbative expansion, Eq.~\eqref{eq_vacuum_mass_phi4_perturbative}, and the classical relation $b(g) = \sqrt{24 g}$, i.e. with $M_{\text{sh-G}}(\sqrt{24 g})$. Both these estimates are qualitatively and quantitative worse than the variational approximation of the mass, $M_{\text{sh-G}}(b(g))$, in the range $g \leq 1/3$.

Remarkably, although the computation is performed within a different theory, the resulting variationally-approximated mass closely tracks that of the Borel-resummed $\varphi^4$ mass for couplings $g \lesssim 1/3$, providing a compelling indication of the robustness and internal consistency of the variational method. At the same time, the variational approximation fails to reproduce the vanishing of the physical mass at the Chang duality point. Within our construction, which employs the sinh–Gordon mass as the sole dynamical input, the only signature of criticality arises at the limiting value $b^\star$, corresponding to $\gmax$, which is approached only asymptotically by the function $b(g)$.

\subsection{A caveat for Hartree-like Approximations}
The relation in Eq.~\eqref{eq_formal_g_lambda} determines the coupling $b(g)$ by minimizing the vacuum energy of the $\varphi^4$ theory with respect to that of the sinh–Gordon model. It is therefore natural to ask whether a Hartree-like approximation may also be employed to estimate other observables of the $\varphi^4$ theory, using the known results for a free field theory with an appropriately defined effective mass. In this approach—see Appendix A of \cite{Chang:1976ek}—the $\varphi^4$ Hamiltonian density Eq.~\eqref{eq_phi4_hamiltonian}, normal ordered with respect to the same mass $m$, is rewritten by splitting the field $\varphi$ into a free component of mass $m'$ and a constant shift. In conventional treatments, this constant is fixed self-consistently through the vacuum expectation values (VEVs) of the field powers. In our framework, however, natural candidates for these quantities are provided by the sinh–Gordon VEVs, Eq.~\eqref{eq_vev_powers_shG}.

Since $\braket{\varphi}_{\text{sh-G}} = 0$ and $\braket{\NO{\varphi^2}}_{\text{sh-G}}$ is strictly positive, we introduce the decomposition
\begin{equation}\label{eq_hartree_splitting}
\varphi = \pm \sqrt{\braket{\NO{\varphi^2}}_{b(g)}} + \phi{m'},
\end{equation}
where $\braket{\NO{\varphi^2}}_{b(g)}$ is evaluated through the relation $b(g)$ in Eq.~\eqref{eq_formal_g_lambda}.
The idea underlying this Hartree-like approximation is to determine $m'$, for given $g$ and $m$, via a variational condition: the operator is again the $\varphi^4$ Hamiltonian density, while the trial state is chosen as the vacuum of the free field of mass $m'$. The self-consistency requirement yields
\begin{equation}\label{eq_hartree_mass}
\frac{m'^2}{m^2} + \frac{3g}{\pi}\ln\frac{m'^2}{m^2}
= 1 + 12 g\braket{\NO{\varphi^2}}_{b(g)}.
\end{equation}
Neglecting the logarithmic correction reproduces the standard Hartree mass shift, obtained by replacing the quartic interaction in the Hamiltonian density with the effective term
$6\;\braket{\NO{\varphi^2}}_{b(g)}\;\varphi^2$. As expected, when $g \to 0$, Eq.~\eqref{eq_hartree_mass} yields the trivial limit $m' = m$, corresponding to the free-field theory.
However, because $\braket{\NO{\varphi^2}}_{b(g)}$ is positive definite, the ratio $m'/m$ increases with $g$ and approaches $12 g\braket{\NO{\varphi^2}}_{b(g)}$ at strong coupling. This immediately exposes a limitation of the Hartree–Fock approximation within our variational framework: whereas the physical mass of the $\varphi^4$ excitation decreases monotonically with increasing $g$, the Hartree effective mass behaves oppositely.

We therefore conclude that the Hartree-like approximation, when the self-consistency condition is replaced by Eq.~\eqref{eq_hartree_mass}, fails to provide reliable predictions for the observables of the $\varphi^4$ theory. In essence, our variational construction employs trial states that are not Gaussian, and thus incompatible with the assumptions underlying the Hartree–Fock method.

\section{Variational Method at a Finite Volume: Vacuum Expectation Values and Matrix Elements}\label{s_TBA}

Our goal here is to compute the ground state of the $\varphi^4$ theory on the cylinder of finite circumference $R$.

We again split the Hamiltonian of $\varphi^4$ into $\mathscr H_{\text{sh-G}}$ part and the remainder
\begin{equation}
\label{E-fin-R}
E_{\varphi^4}(R)=E_{\text{sh-G}}(R)+\langle 0|:V:_{M,R}|0\rangle,
\end{equation}
here
\begin{equation}
V=\frac{m^2}{2}\varphi^2+m^2 g\varphi^4-2\mu\cosh(b\varphi),
\end{equation}
with $\mu = \left[m^2/(16\pi b^2)\right] \left(m e^{\gamma_E}/2\right)^{2b^2}$. Here the value of $b$ in the sinh-Gordon model is already fixed as $b=b^*(g)$ from the variational principle for the bulk energy, see Section \ref{s_variational}. 

The first term in \eqref{E-fin-R} is the energy of the sinh-Gordon model with $b=b^*(g)$ on the cylinder, it can be computed using the Thermodynamic Bethe ansatz (TBA), see \cite{Zamolodchikov:1989cf}. The second term can be computed using the special form factor series described below.

\subsection{Thermodynamic Bethe Ansatz \label{Bethe-thermal}}

The energy of $\mathcal H_{\text{sh-G}}$ on the cylinder is given by
\begin{equation}
\label{EShG}
E_{\text{sh-G}}(R)=R\:\mathcal E_{\text{sh-G}}-M_{\text{sh-G}}\int_{-\infty}^{\infty}\frac{du}{2\pi}\cosh u\;\log\left(1+e^{-\varepsilon(u)}\right),
\end{equation}
with the energy density on the plane ($R=\infty$) $\mathcal E_{\text{sh-G}}$ is given by \eqref{eq_vacuum_energy_density_shG_divergent}. Here
the pseudoenergy $\varepsilon(u)$ is determined by the TBA \cite{Zamolodchikov:1989cf} equation 
\begin{equation}
\label{TBA}
\varepsilon(\theta)=M_{\text{sh-G}}R\cosh\theta-\int_{-\infty}^{\infty}\frac{du}{2\pi}\delta(\theta-u)\log\left(1+e^{-\varepsilon(u)}\right).
\end{equation}
Here $\delta$ is given by the logarithmic derivative of $S$-matrix
\begin{equation}
\delta(\theta)=-i\frac{d}{d\theta}\log S(\theta).
\end{equation}

\subsection{Finite volume energy via form factor series\label{cylinder}}

In order to compute the expectation value $\langle\dots\rangle$ of the operator $O$ on the cylinder of a finite radius $R$, starting from the theory on a plane, one can apply the LeClair-Mussardo (LM) formula \cite{Leclair:1999ys}
\begin{equation}
\label{LM-series}
\langle O\rangle_R=\sum_{n=0}^{\infty}\frac{1}{n!}
\left(\prod_{i=1}^n \int\frac{d\theta_i}{2\pi}f(\theta_i)\right)F_{n,c}^O(\theta_1,\dots,\theta_n),
\end{equation}
where 
$\varepsilon(\theta)$ is the energy of one particle with the rapidity $\theta$ (it is determined by \eqref{TBA}),
\begin{equation}
\label{fermi_weight}
f(\theta)=\frac{1}{1+e^{\varepsilon(\theta)}}
\end{equation}
and $n$-particles {\it connected form factors} $F_n^O(\theta_1,\dots,\theta_n)$ with particle rapidities $\{\theta_1,\dots,\theta_n\}$ are defined by
\begin{equation}
\label{FF_connected}
F_{n,c}^O(\theta_1,\dots,\theta_n)=\text{FP}\lim_{\eta_i\to0}\langle\theta_n,\dots,\theta_1|O|\theta_1+\eta_1,\dots,\theta_n+\eta_n\rangle,
\end{equation}
where $\text{FP}$ stands for {\it finite part}. $\text{FP}$ is defined in a way that any term proportional to $\eta_i^{-k}$, $k>0$ and any term proportional to $\eta_i/\eta_j$ $\forall\; i\ne j$ is discarded. Here, the computation of connected form factors is performed on a plane ($R=\infty$). The first term in \eqref{LM-series} corresponds to computation on an infinite plane, since for $n=0$ connected form factors coincide with the ordinary one.

We can use \eqref{LM-series} for the computation of the energy of the $\varphi^4$ theory as well, considering $O=\mathscr H_{\varphi^4}$ as an operator. Of course, the $\varphi^4$ model is not integrable. However, in our computations, we simply take the average of $\mathscr H_{\varphi^4}$ as an arbitrary operator with respect to the vacuum $|0_b\rangle$ of the sh-G model, in the same way as in the case of an infinite system. $\mathscr H_{\varphi^4}$ is an ordinary operator there. Thus, we can apply \eqref{LM-series} to this problem as well.

Thus, combining \eqref{E-fin-R}, \eqref{EShG} and the results of Section \ref{Bethe-thermal} and then subtracting term $\mathcal E_{\text{sh-G}}R$, which corresponds to that bulk energy, we obtain the contribution that explicitly defines the energy on the cylinder
\begin{equation}
\label{energy-R}
E_{\varphi^4}(R)=-M_{\text{sh-G}}\int_{-\infty}^{\infty}\frac{du}{2\pi}\cosh u\;\log\left(1+e^{-\varepsilon(u)}\right)+\sum_{n=1}^{\infty}\frac{1}{n!}\left(\prod_{i=1}^n\int_{-\infty}^{\infty}\frac{d\theta_i}{2\pi} f(\theta_i)\right)F_{n,c}^{:V:}(\theta_1,\dots,\theta_n).
\end{equation}
Hence, it remains to compute the contributions explicitly coming from the series \eqref{LM-series}.

Form factors of the exponential operator were computed in \cite{Fring:1992pt,Mussardo:1993ut,koubek1993operator}, required results can be found in Appendix \ref{a_TSM_explanation} and connected form factors can be computed by taking an appropriate limit \eqref{FF_connected} \cite{Kormos:2009yp}. Explicitly, the first few connected form factors are given by
\begin{equation}\label{elementary_FF_formula}
\langle0|:e^{a\varphi}:|\theta_1,\dots,\theta_n\rangle=\langle e^{b\varphi(0)}\rangle[a]\left(\frac{4\sin(\pi \alpha)}{F_\text{min}(i\pi)}\right)^{\frac{n}{2}}Q_n(a)
\prod_{i<j}^n\frac{F_{\text{min}}(\theta_i-\theta_j)}{x_i+x_j}.
\end{equation}
Here
\begin{equation}
\label{fmin}
F_{\text{min}}(\theta)=\mathcal 
N\exp\left\{8\int_0^{\infty}\frac{dt}{t}\frac{\sinh\left(\frac{t}{2}\alpha\right)\sinh\left(\frac{t}{2}(1-\alpha)\right)\sin\left(\frac{t}{2}\right)}{\sinh^2(t)}\sin^2\left(\frac{t\hat\theta}{2\pi}\right)\right\},
\end{equation}
\begin{equation}
\label{norm}
\mathcal N=F_{\text{min}}(i\pi)=\exp\left\{-4\int_0^{\infty}\frac{dt}{t}\frac{\sinh\left(\frac{t}{2}\alpha\right)\sinh\left(\frac{t}{2}(1-\alpha)\right)\sinh\left(\frac{t}{2}\right)}{\sinh^2(t)}\right\}.
\end{equation}
Above we have used the following notation: $\hat\theta=i\pi-\theta$ and $x_i=\exp(\theta_i)$, together with 
\begin{equation}
[k]\equiv\frac{\sin(k\pi\alpha)}{\sin(\pi\alpha)}.
\end{equation}
The polynomials $Q_n$ are given by $Q_n(k)=\det M_n(k)$
\begin{equation}
\label{Q-det}
[M_n(k)]_{ij}=\sigma^{(n)}_{2i-j}\times[i-j+k],
\end{equation}
where $\sigma^{(n)}_k$ are the elementary symmetric polynomials defined in terms of the generating function 
\begin{equation}
\label{sigma-generator}
\prod_{i=1}^n(x+x_i)=\sum_{k=0}^n x^{n-k}\sigma^{(n)}_k(x_1,\dots,x_n).
\end{equation}

Correspondingly, the connected form factors of the exponential operators can be derived using the limit \eqref{FF_connected} from \eqref{elementary_FF_formula} and are given by
\begin{equation}
\label{connected-exp}
\begin{split}
&F_{1,c}^{e^{a\varphi}}(\theta_1)=\langle \theta_1 |:e^{a\varphi}:| \theta_1\rangle=\frac{4}{\mathcal N}\frac{\sin^2(a\pi\alpha)}{\sin(\pi \alpha)}\langle :e^{a\varphi}: \rangle,\\
&F_{2,c}^{e^{a\varphi}}(\theta_1,\theta_2)\\
&=\langle \theta_1,\theta_2 |:e^{a\varphi}:|\theta_1,\theta_2\rangle=\frac{16}{\mathcal N}\frac{[a]\sin^2(\pi \alpha)}{\sinh^2(\theta_1-\theta_2)}\left(\cosh(\theta_1-\theta_2)[a]^2-[a+1][a-1]\right)\langle :e^{a\varphi}: \rangle,
\end{split}
\end{equation}
and therefore the connected form factors of $\varphi^2$ and $\varphi^4$ are given by the expansions of \eqref{connected-exp} in the Maclaurin series, i.e. 
\begin{equation}
\begin{split}
F^{:\varphi^2:}_{1,c}(\theta_1)=\left.\frac{\partial^2}{\partial a^2} F^{e^{a\varphi}}_{1,c}(\theta_1)\right|_{a=0},\qquad
F^{:\varphi^4:}_{2,c}(\theta_1,\theta_2)=\left.\frac{\partial^4}{\partial a^4} F^{e^{a\varphi}}_{2,c}(\theta_1,\theta_2)\right|_{a=0}.
\end{split}
\end{equation}
Substituting \eqref{connected-exp} in a series \eqref{energy-R} and performing numerical integration, we obtain finite radius contributions to the energy of the $\varphi^4$ model.

Higher contributions can be computed as well; however, it is easy to see that the $(n+1)$-th term in the series \eqref{LM-series} is suppressed in comparison to the $n$th term as $o(e^{-MR})$. Thus, at $MR\ge 1$ the series converges rapidly, and the first few contributions dominate.

\begin{figure*}
\centering
\begin{subfigure}{0.49\linewidth}
\includegraphics[width=\linewidth]{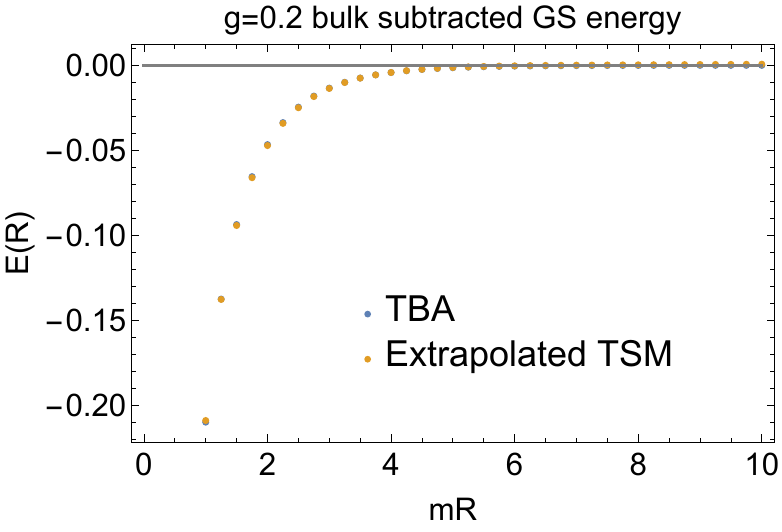}
\subcaption{}
\end{subfigure}
\hfill
\begin{subfigure}{0.49\linewidth}
\includegraphics[width=\linewidth]{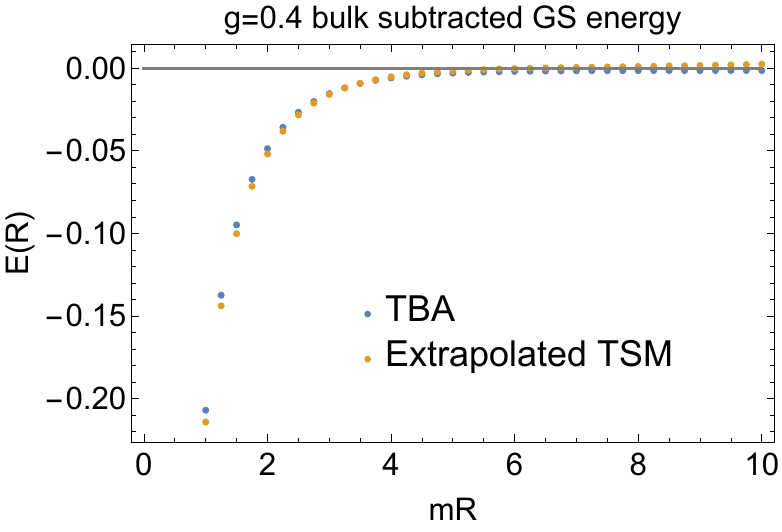}
\subcaption{}
\end{subfigure}%
\caption{a) Ground state energy of $\mathscr H_{\varphi^4}$ with $g=0.2$ on the cylinder (blue dots) computed using the combination of the TBA and the LeClair-Mussardo series. The results \st{is} are compared with \st{the} results coming from the free massive boson-based Hamiltonian truncation (see section \ref{s_TSM}), shown as the orange dots. Bulk energy \eqref{eq_vacuum_energy_density_shG_divergent} is subtracted.  b) The same for $g=0.4$.}
\label{TBA+LM-TSM}
\end{figure*}

The numerical comparison between Eq.~\eqref{energy-R} and the results obtained from the Truncated Space Method (TSM) (see Section~\ref{s_TSM}) is presented in Fig.~\ref{TBA+LM-TSM}. Remarkably, the finite-volume energy of the $\varphi^4$ model on the cylinder, as computed within the TSM framework, exhibits a surprisingly close agreement with that of the sinh–Gordon model evaluated through the TBA. The second contribution in Eq.~\eqref{energy-R}, arising from the LeClair–Mussardo (LM) series, accounts for only a few percent correction numerically.
The quality of this correspondence in $E(R)$ is particularly striking when one recalls that the bulk energy density $\mathcal{E}$ of the $\varphi^4$ model—computed using the sinh–Gordon basis—shows a substantial deviation from the exact bulk energy of the $\varphi^4$ theory at large coupling, as illustrated in Fig.~\ref{f_energy_mass}a.

\section{Finite Volume Variational Method through Truncated Space Approach}\label{s_TSM}
In this section, we present the exact diagonalization of the Truncated Space Method (TSM) Hamiltonian for the $\varphi^4$ theory. In contrast to previous studies, which employed the free-boson basis \cite{Lassig:1990xy, Delfino:1996xp, Fitzpatrick:2022dwq, 2015NuPhB.899..547K, 2021JHEP...01..014K, Coser:2014lla, konikReviewTSA, LajerKonik, Horvath:2022zwx, Takacs2012, BERIA2013457, BAJNOK201493,Xu:2022mmw,Xu:2023nke,Brandino:2010sv}, we adopt as our computational basis the interacting sinh–Gordon model, thereby incorporating nontrivial dynamical correlations from the outset. The idea of this method was outlined in the context of sinh-Gordon to sinh-Gordon perturbations in \cite{2021JHEP...01..014K}. It is applied for the first time to a nonintegrable model in the present work. The details of the technical implementation\footnote{We plan to make our FFTSM implementation publicly available on GitHub upon publication of this work; in the meantime, the code can be provided upon reasonable request.} are provided in Appendix~\ref{a_TSM_explanation}.

We then carry out a systematic analysis of the raw TSM spectra for various values of the couplings $g$ and $b$. Our investigation focuses on the regime $g \leq 1/3$, which—as shown in Fig.~\ref{f_energy_mass}—corresponds to the range where the variational predictions are in quantitative agreement with the Borel-resummed perturbative series, Eq.~\eqref{eq_vacuum_mass_phi4_perturbative}. In this domain, the maximum deviation between the infinite-volume variational result and the resummed perturbation theory does not exceed 10$\%$.

In what follows, we will analyze the finite-size effects arising both from the finite spatial extent $mR$ and from the finite truncation dimension of the Hilbert space, $d_{\text{TSM}}$, in order to assess the convergence and reliability of the TSM within this framework.

\subsection{Numerical Setup}

As in standard perturbative analyses, we impose cutoffs both on the particle number and on the maximum single-particle momentum. Specifically, each TSM basis state contains at most six particles, and each particle momentum is restricted to $|p| \leq 18$ (in units of the boson mass $m = 1$). Physically, this truncation neglects elastic and inelastic processes involving more than six particles, which have a negligible contribution to the observables considered.

In the large-volume limit, only a few low-energy sinh–Gordon states contribute significantly to the low-lying spectrum of the $\varphi^4$ model, whereas, at smaller volumes, higher excited states can play a more prominent role. To control this behavior, we introduce a dimensionless cutoff $N_c$ \cite{konikReviewTSA}, which defines an energy cutoff
\begin{equation}
E_c = \frac{2\pi N_c}{R},
\end{equation}
where $R$ denotes the spatial circumference of the system. The resulting Hilbert-space dimensions for various values of $R$ are reported in Table~\ref{t_dim_TSM}.
The choice of $N_c$—which effectively reduces the basis size as the system radius increases—is particularly advantageous for exploring a broad range of volumes, $3 \leq R \leq 16$. Indeed, the computational bottleneck of the interacting TSM lies in the evaluation of off-diagonal matrix elements, which must be computed from the finite-size form-factor expressions and scale rapidly with the basis dimension.
\begin{table}[]
    \centering
    \begin{tabular}{|l|l|l|l|l|l|}
    \hline
        $N_c$ & $R=1$ & $R=4$ & $R=8$ & $R=16$ & $R=24$ \\ \hline
        0 & 1 & 1 & 1 & 1 & 1 \\ \hline
        2 & 4 & 2 & 1 & 1 & 1 \\ \hline
        4 & 7 & 6 & 3 & 1 & 1 \\ \hline
        6 & 16 & 12 & 6 & 3 & 1 \\ \hline
        8 & 32 & 29 & 13 & 5 & 3 \\ \hline
        10 & 66 & 61 & 31 & 6 & 5 \\ \hline
        12 & 113 & 108 & 64 & 14 & 6 \\ \hline
        14 & 203 & 196 & 123 & 30 & 7 \\ \hline
        16 & 320 & 313 & 215 & 50 & 12 \\ \hline
        18 & 512 & 503 & 355 & 98 & 30 \\ \hline
        20 & 766 & 757 & 578 & 188 & 52 \\ \hline
        22 & 1137 & 1126 & 902 & 325 & 84 \\ \hline
        24 & 1602 & 1591 & 1334 & 536 & 138 \\ \hline
        26 & 2273 & 2264 & 1936 & 867 & 234 \\ \hline
        28 & 3080 & 3073 & 2717 & 1323 & 407 \\ \hline
        30 & 4167 & 4160 & 3719 & 1969 & 680 \\ \hline
        32 & 5491 & 5484 & 4993 & 2793 & 1082 \\ \hline
        34 & 7201 & 7194 & 6634 & 3887 & 1665 \\ \hline
        36 & 9216 & 9209 & 8611 & 5308 & 2474 \\ \hline
        38 & 11794 & 11787 & 11094 & 7141 & 3545 \\ \hline
    \end{tabular}
    \caption{Dimensions of the TSM Hilbert space, varying the adimensional cut-off $N_c$, for various volumes $R$. A cut-off $N_c=8$ provides in the range of volumes $1 \leq R \leq 16$ with at least three two-particle states with nonzero individual momenta. These will be used to fit the $S$-matrix phase below threshold.}
    \label{t_dim_TSM}
\end{table}

\subsection{Variational Curve from Overlap Maximization}

It is well known that, in the large-$R$ limit, the finite-volume corrections to the lowest energy states become exponentially suppressed. This allows us to isolate and analyze the finite-size effects arising solely from the finite dimensionality of the TSM basis.

To this end, we quantify the contribution of the sinh–Gordon ground state $\ket{0_b}$ to the $\varphi^4$ TSM ground state $\ket{E_0(g;R)}$. Fixing $R = 8$ and $N_c = 8$, we examine the overlap
\begin{equation}
\mathscr{P}_b(g; R) = 1 - \big|\braket{0_b | E_0(g; R)}\big|^2,
\end{equation}
which measures the total weight of all sinh–Gordon basis states other than $\ket{0_b}$ contributing to $\ket{E_0(g;R)}$ at a given coupling $g$, as $b$ is varied. In addition to the vacuum, the $\varphi^4$ ground state receives contributions from twelve even-particle sinh–Gordon states. We thus identify the values of $b$ that minimize $\mathscr{P}_b(g;R)$, providing an independent variational estimate complementary to that obtained from the minimization of the ground-state energy.

\begin{figure*}
\centering
\begin{subfigure}{0.48\linewidth}
\includegraphics[width=\linewidth]{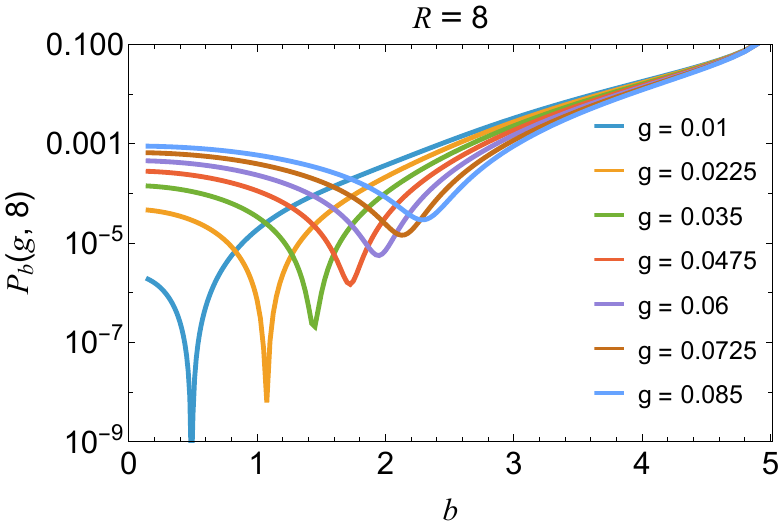}
\subcaption{}
\end{subfigure}
\hfill
\begin{subfigure}{0.48\linewidth}
\includegraphics[width=\linewidth]{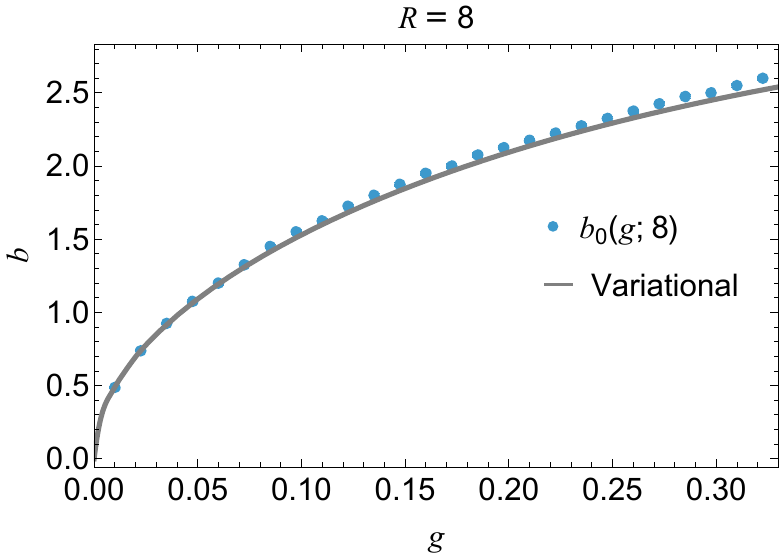}
\subcaption{}
\end{subfigure}%
\caption{a) Plot of the overlap $\mathscr{P}_b(g;R)$ for $R=8$, $N_c=8$, as function of the sinh-Gordon coupling $b$, varying the $\varphi^4$ coupling $g$. Each individual curve presents nontrivial minima $b_o (g;R)$. b) Plot of the minima $b_o(g;R)$ (blue dots), with numerical uncertainties, varying $g$. They line agrees with the optimal curve Eq.~\eqref{eq_formal_g_lambda} (here the gray solid curve). The deviations are due to the finite volume and we checked that increasing the volume, the curve collapses to the variational one.}
\label{f_overlap}
\end{figure*}

We explore sample couplings in the range $0.01 \leq g \leq 1/3$, and study the overlap $\mathscr{P}_b(g;R)$ for $0.15 \leq b \leq b_\star$, as shown in Fig.~\ref{f_overlap}(a). The overlap exhibits a non-monotonic dependence on $b$, displaying distinct minima at $b = b_o(g;R)$. Remarkably, as illustrated in Fig.~\ref{f_overlap}(b), the locations of these minima closely coincide with the infinite-volume variational curve obtained from energy minimization.

At finite volume, the overlap criterion and the variational-energy criterion are related: for the trial state $|0_b\rangle$, one has
\[
\langle 0_b|H_{\phi^4}(R)|0_b\rangle - E_0(R)
= \Delta_b(R)\Bigl(1-|\langle E_0(g;R)|0_b\rangle|^2\Bigr),
\]
where $\Delta_b(R)$ denotes the average excitation energy of the component orthogonal to the ground state,
\begin{equation}
\Delta_b(R)=\sum_{n>0}|\langle n|0_b\rangle|^2(E_n-E_0)/\left(\sum_{n>0}|\langle n|0_b\rangle|^2\right).
\end{equation}
Therefore, one expects the minima of $\mathcal{P}_b(g;R)$ to track the variational curve, although not to coincide exactly, since $\Delta_b(R)$ is in general both $b$- and volume-dependent. Physically, our result implies that the optimal coupling $b$ that minimizes the vacuum energy simultaneously maximizes the overlap between the $\varphi^4$ ground state and the sinh–Gordon vacuum to a good approximation -- a powerful indication of the consistency and coherence of the variational construction.

\subsection{Infinite-Energy Extrapolation}

The goal of this section is to study the infinite-energy extrapolation properties of the TSM. A practical way is to generate a very large basis, with some large number of particles and maximum particle momentum, such that the maximum energy is $E_{\text{max}}$. Then we consider only the states with energy below a cut-off threshold $E_c$ and construct the TSM Hamiltonian. the infinite-energy extrapolation is the study of the same particle line (vacuum, one-particle and higher-energy lines) obtained from the TSM Hamiltonians varying the cut-off $E_c$ \cite{Rychkov:2014eea}.

We focus on the representative case $g = 0.1$, with $b = b(g) = 0.489686\ldots$ and a system size $L = 8$. The chosen volume is sufficiently large to probe the cutoff dependence while suppressing finite-volume effects. After diagonalizing the Hamiltonian for successive values of $E_c$, we track each individual energy level and fit it using the extrapolation formula
\begin{equation}
E_i = E_i^\infty + b_1 \frac{\log^2 E_c}{E_c^2} + b_2 \frac{\log E_c}{E_c^2}.
\end{equation}
Even for relatively modest cutoffs ($E_c \leq 13$), we observe a systematic convergence of the extrapolated ground-state energy $E_0^\infty$. The corresponding energy density, $-5.04708(4) \times 10^{-6}$, lies intermediate between the Borel-resummed value at the same coupling, $-5.0478 \times 10^{-6}$, and the infinite-volume variational prediction, $-5.04593 \times 10^{-6}$.

\begin{figure*}
\centering
\begin{subfigure}{0.48\linewidth}
\includegraphics[width=\linewidth]{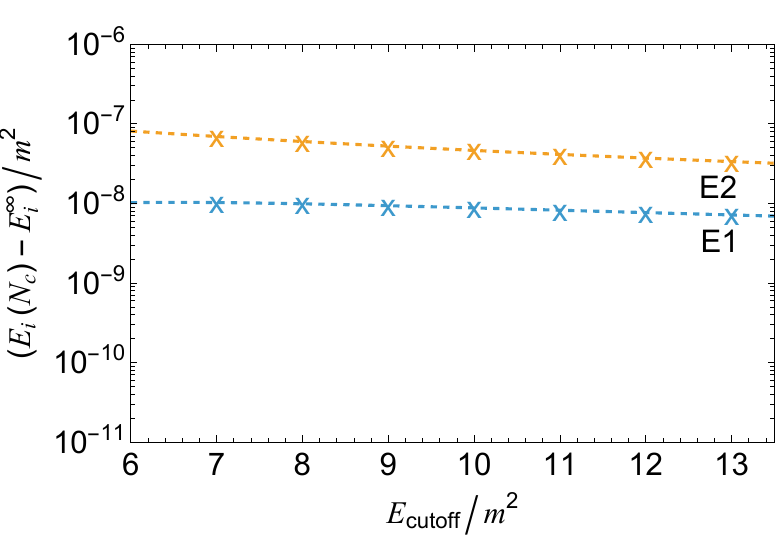}
\subcaption{}
\end{subfigure}
\hfill
\begin{subfigure}{0.48\linewidth}
\includegraphics[width=\linewidth]{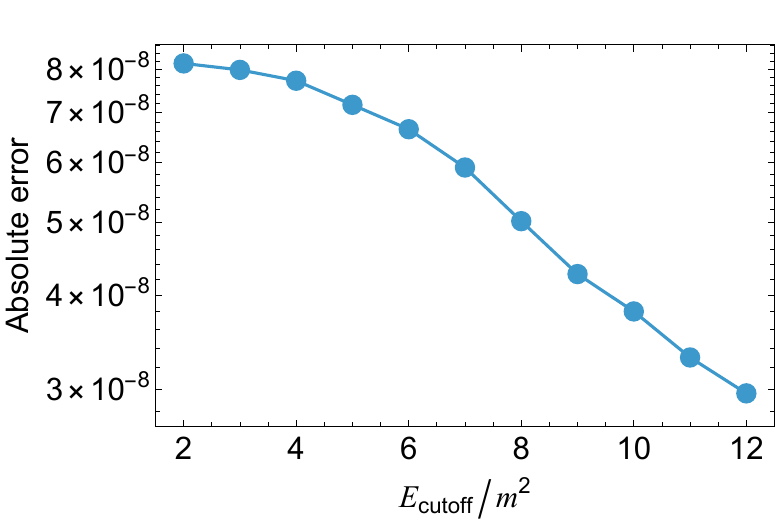}
\subcaption{}
\end{subfigure}%
\caption{a) Extrapolation curves as function of the cut-off for the ground state ($E_1$) and 2-particle state ($E_2$) for $g=0.1$ and $L=8$. The variational principle provides a good ansatz for the optimal coupling $b_g$. This improves significantly the performance of the TSM, which provides very good estimates even with a small basis. b) Absolute error between the first 2-particle line and the ground state, as function of the cut-off. The reference value the one obtained by extrapolation. Even for small cut-offs, the relative error is significantly lower than the magnitude of the bare 2-particle line.}
    \label{f_extrapolation}
\end{figure*}

Although extrapolation to infinite cutoff may be required to obtain highly accurate estimates of individual energy levels, this is not necessarily the case for energy differences. In particular, the gap between the ground state and the first two-particle excitation—relevant for the computation of the elastic $S$-matrix below the inelastic threshold—is subject to a benign cancellation of cutoff effects, as illustrated in Fig.~\ref{f_extrapolation}.

From this extrapolation analysis, we observe that the raw numerical values, obtained without any additional post-processing, are only marginally affected by finite-size effects. We attribute this remarkable stability to the choice of basis, which inherently incorporates the finite-volume form factors in both the diagonal and off-diagonal matrix elements of the Hamiltonian. This outcome provides strong evidence for the variational consistency and numerical robustness of the sinh–Gordon basis, confirming its effectiveness in capturing the essential finite-volume dynamics of the $\varphi^4$ theory.

\subsection{Optimal relation $b_g(R)$ at Finite Volume}
\begin{figure*}
\centering
\begin{subfigure}{0.48\linewidth}
\includegraphics[width=\linewidth]{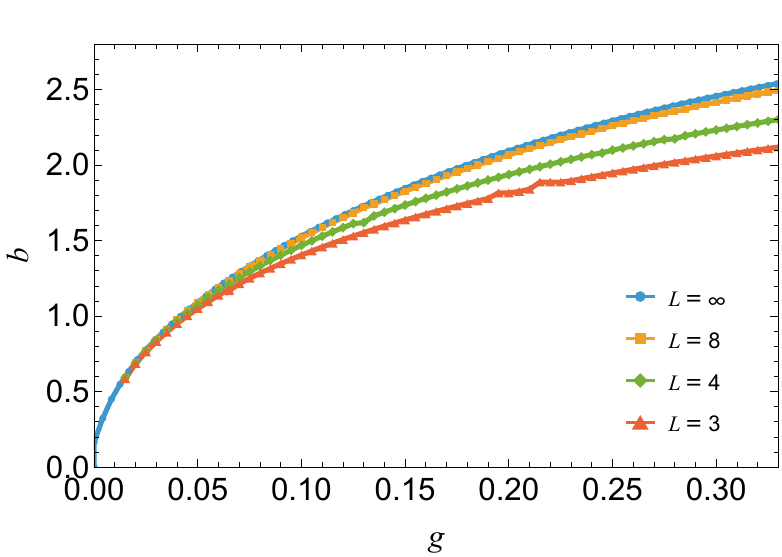}
\subcaption{}
\end{subfigure}
\hfill
\begin{subfigure}{0.48\linewidth}
\includegraphics[width=\linewidth]{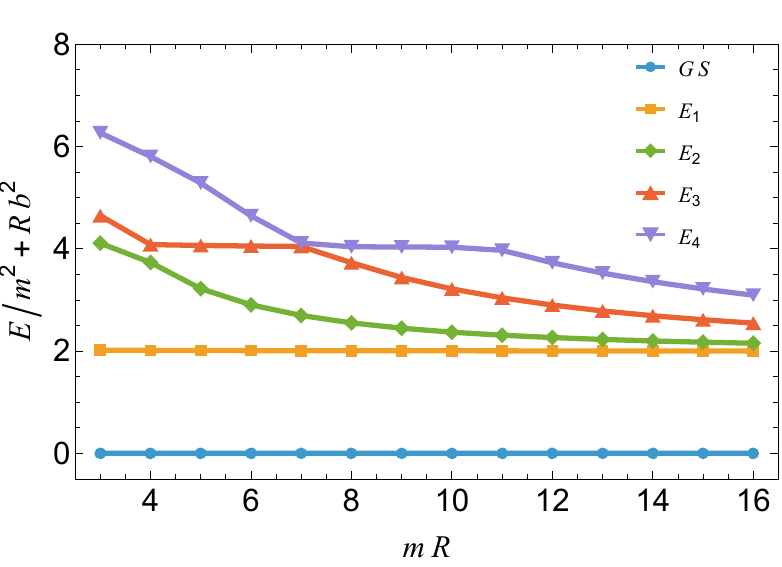}
\subcaption{}
\end{subfigure}%
\caption{a) Optimal curve relating the truncated $\varphi^4$ Hamiltonian and the sinh-Gordon as function of the volume $3\leq R\leq 16$. For sufficiently large volumes the $b_g(R)$ curve collapse to the infinite-volume one (blue). For smaller volumes, the curve grows slower. Still, deviations are not extreme, and this makes the variational principle reliable not only at infinite volumes but also at finite ones. b) Plot of the first low-lying energies of the parity-even sector of the truncated $\varphi^4$ Hamiltonian, without the divergent contribution $-R m^2/b^2$, for fixed $g=0.05$. Because of this removal, the characteristic linear decrease at large volumes is not visible in the plot, but is present. Higher-particle states with nonzero individual momentum feature level repulsion as the Hamiltonian is not integrable.}
\label{f_TSM_optimal}
\end{figure*}

With the cutoff fixed at $N_c = 8$, we examine the behavior of the optimal coupling curve $b_g(R)$ as a function of the system size in the range $3 \leq R \leq 16$. For each pair of parameters $(R, g)$, the ground-state energy of the truncated $\varphi^4$ Hamiltonian is minimized with respect to $b$. The resulting curves $b_g(R)$ for selected volumes, together with the corresponding low-lying energy spectra in the parity-even sector, are displayed in Fig.~\ref{f_TSM_optimal}.

For sufficiently small couplings, the perturbative quadratic relation $b \sim g^{1/2}$ remains largely unaltered by finite-volume effects. However, as the volume decreases, the system becomes increasingly sensitive to the limited size of the interacting basis, and deviations from the infinite-volume behavior emerge, reflecting the progressive impact of finite-volume corrections on the variational landscape.

\subsection{Estimate of the $S$-matrix Phase Below Threshold}

To estimate the $S$-matrix phase shift below the four-particle threshold, we employ the Bethe–Yang quantization condition.
If $E$ denotes the energy of a two-particle state with particle mass $M$, the corresponding rapidity $\theta$ is obtained by inverting
$E = 2M \cosh\theta$, with the momentum of the particle given by:
\begin{equation}
    p = M \sqrt{\left(\frac{E}{2M}\right)^2-1}.
\end{equation}
A two-particle state in the zero momentum sector may thus be represented as $\ket{-\theta, \theta}$. The asymptotic incoming and outgoing states coincide up to an overall $S$-matrix phase,
\begin{equation}
S(2\theta) = \exp\left(i \delta(2\theta) \right)\,\,\,.
\end{equation}
Because of periodic boundary conditions, the same state is also obtained by translating one of the particles by the spatial circumference $R$, leading to the Bethe–Yang quantization equation
\begin{equation}
\delta(2\theta) + M R \sinh\theta = 2\pi n,\label{BetheYang2pt}
\end{equation}
where $n$ is an integer (or half-integer, depending on the parity sector).

Below the inelastic threshold, the $\varphi^4$ $S$-matrix can be conveniently parameterized in terms of a CDD factor $\alpha$ as
\begin{equation}
\delta(\theta; \alpha)
= -\frac{\pi}{2}
+ \arctan\left[
\frac{\sinh^2\theta - \sin^2\pi\alpha}{
2\sinh\theta \sin\pi\alpha}
\right].
\end{equation}

This parametrization will later allow us to contrast the extracted phase shifts with the variationally determined coupling $b_g$.

\begin{figure*}
\centering
\begin{subfigure}{0.48\linewidth}
\includegraphics[width=\linewidth]{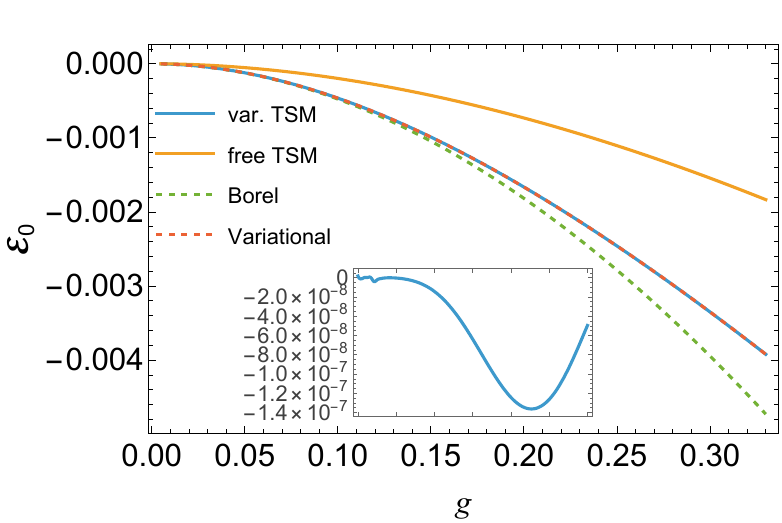}
\subcaption{}
\end{subfigure}
\hfill
\begin{subfigure}{0.48\linewidth}
\includegraphics[width=\linewidth]{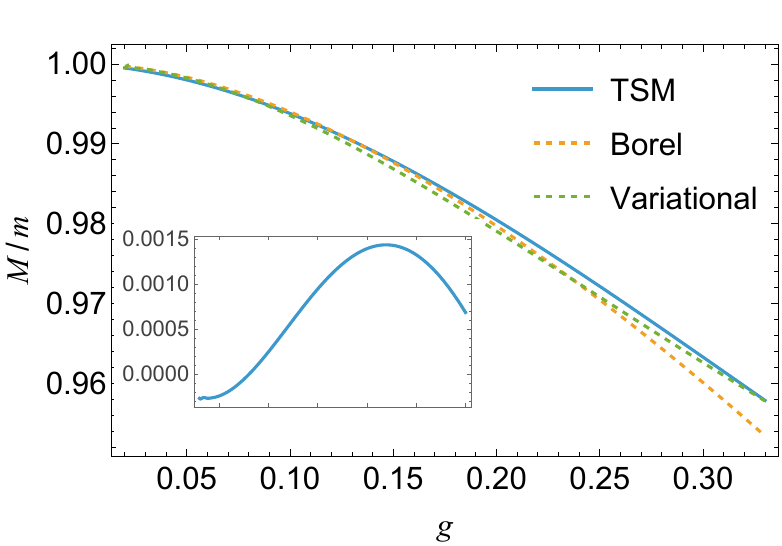}
\subcaption{}
\end{subfigure}%
\caption{a) Estimate of the ground state energy density of $\varphi^4$ using the variational TSM, compared with the Borel resummed curve and the variational one. The inset shows the difference between the TSM and variational curves. For the range of parameters chosen, $g\leq 1/3$, the TSM curve overlaps with the variational one. The results from the free bosonic TSM are shown for a basis consisting of 8 zero modes and 125711 massive states. Comparing to the variational TSM, it is clear that the choice of the integrable basis, already without any energy cut-off extrapolation, provides a better initial value of $\mathcal{E}_0$. Therefore the variational TSM already provides better estimates with a much smaller basis size, compared to the free boson implementation.  b) However, the mass estimated through the TSM overlaps much better with the Borel resummed curve rather then the variational one (as before, the inset shows the difference between the TSM and variational curves). In fact, the variational one is simply $M_\text{sh-G}(b_g)$, while the TSM is obtained directly from the $\varphi^4$ Hamiltonian at finite volume, henceforth it should match better the Borel resummed curve.}
\label{f_TSM_energy_density_mass}
\end{figure*}

A key ingredient in the numerical extraction of the phase shift is the mass of the $\varphi^4$ excitation, which can be determined through the following procedure:

\begin{itemize}
    \item 
    {\bf Ground-state energy density.}
The vacuum energy density $\mathcal{E}(g)$ is obtained by fitting the ground-state energy as a function of the system size $R$.
     \item {\bf One-particle mass.}
For the largest available volume, $R_{>}$, the spectrum is computed in the odd-parity sector. The corresponding ground state, $E_0^{(o)}(R_{>}, g, b_g)$, yields the one-particle energy
    \begin{equation}
    E_{1 p}(R_{>},g,b_g) \,=\, E_0^{(o)}(R_{>}, g, b_g) - R_{>} \,{\mathcal E}(g) \equiv M_g \,\,\,.
    \end{equation}
    \item {\bf Two-particle states.}
In the even-parity sector, each bare eigenvalue corresponding to a two-particle configuration, $E^{(2)}(R, g, b_g(R))$—distinguished by the associated momenta—defines a two-particle energy level. The particular choice of an interacting sinh–Gordon basis appears to significantly suppress finite-size corrections, allowing us to estimate the two-particle energy as
\begin{equation}
E_{2p}(R, g, b_g(R)) =
E^{(2)}(R, g, b_g(R)) - R\,\mathcal{E}(g).
\end{equation}
    \end{itemize}
The resulting ground-state energy densities and masses obtained from the first two steps are displayed in Fig.~\ref{f_TSM_energy_density_mass}. The vacuum energy $\mathcal{E}(g)$ shows excellent agreement with the variational prediction across the studied range $g \leq 1/3$, while the extracted mass $M_g$ follows more closely the Borel-resummed result for $g \lesssim 0.2$, providing a further quantitative validation of the variational framework.

\begin{figure*}
\centering
\begin{subfigure}{0.48\linewidth}
\includegraphics[width=\linewidth]{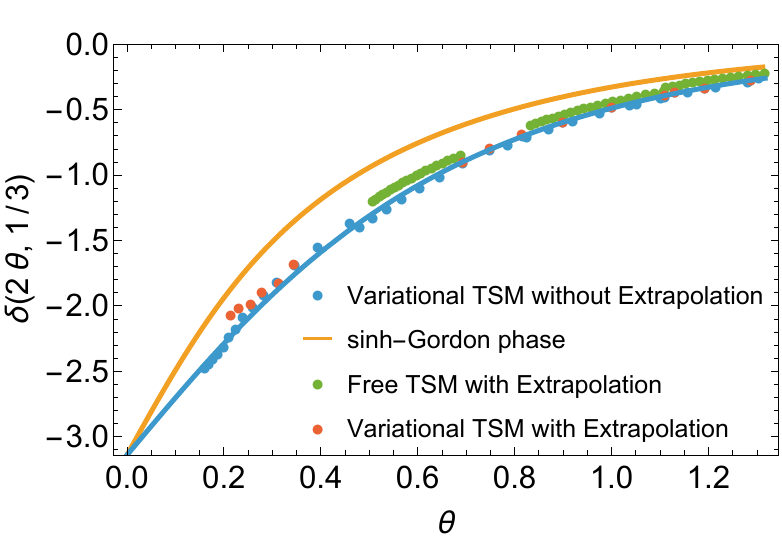}
\subcaption{}
\end{subfigure}
\hfill
\begin{subfigure}{0.48\linewidth}
\includegraphics[width=\linewidth]{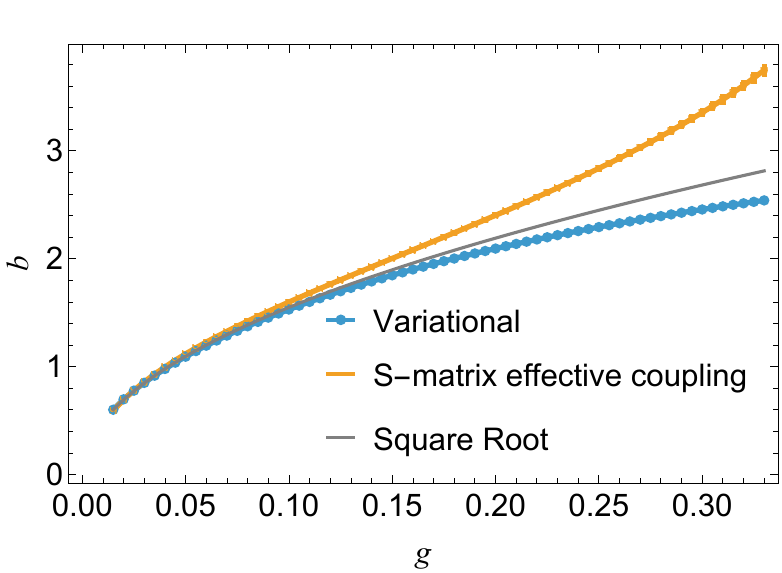}
\subcaption{}
\end{subfigure}%
\caption{a) An example of numerically-determined $S$-matrix phase data for $g=1/3$. We observe clear deviations from the $S$-matrix phase of parameter $b_g$. The best-fit estimate nicely interpolates between the numerical values. Numerical data obtained by infinite-cutoff extrapolation of the free TSM are also shown and are in excellent agreement with the less resourceful variational TSM. b) Plot of the $S$-matrix effective coupling estimate (orange) versus the variational curve and the square root approximation.}
\label{f_TSM_s-matrix}
\end{figure*}

The procedure outlined above enables us to compute the two-particle energy levels for system sizes in the range $3 \leq R \leq 16$.
We extract the phase shift $\delta(2\theta)$ from the two--particle finite--volume spectrum using the Bethe--Yang quantization condition \eqref{BetheYang2pt} in the elastic regime below the four--particle threshold, 
\begin{equation} 
\theta^*=\sinh^{-1}\sqrt{3}\approx 1.31696...
\end{equation}
where inelastic effects are negligible.
For a fixed coupling $g$, the $S$-matrix phases extracted from different two-particle lines show excellent mutual agreement as functions of rapidity, allowing the data to be combined into a single unified dataset. From this dataset, we determine the effective coupling $b'_g$ through a one-parameter fit of the phase shift. An example of this analysis is shown in Fig.~\ref{f_TSM_s-matrix}, which displays both the numerical data points and the corresponding fit for the largest coupling studied, $g = 1/3$, together with the behavior of the effective $S$-matrix coupling as a function of $g$. Across the full coupling range investigated, the numerical data exhibit excellent agreement with the one-parameter fit, with a maximum $\chi^2$ value not exceeding 0.15. The accuracy can be further improved by excluding the two-particle line with zero individual momentum, which predominantly affects the small-rapidity region, where energies are lower. The extracted effective coupling $b'(g)$ shows a noticeable deviation from the variational estimate $b_g$, increasing more rapidly than the expected square-root behavior. This discrepancy indicates that the variational parameter $b_g$, while optimal for the vacuum energy, does not simultaneously optimize the $S$-matrix phase, as the variational approach is not designed to reproduce all observables with equal accuracy.

\paragraph{Comparison with the free-boson TSM.}
To benchmark the efficiency of the present approach against the truncated spectrum method (TSM) built on a free massive boson basis \cite{BajnokMarton2}, we employed the implementation from Ref.~\cite{BajnokMarton2}. We computed the finite-volume eigenvalues using an oscillator basis ranging from $12648$ states (chiral cutoff $12$) down to $435$ states (chiral cutoff $7$), together with a separated zero mode. The zero-mode dynamics were treated exactly by solving the associated $\varphi^4$ quantum mechanics in a $1000$-dimensional Hilbert space; from this, we retained $8$ eigenstates in each $\mathbb{Z}_2$ parity sector (16 in total), and formed the full basis as the tensor product of the zero-mode subspace with the oscillator states. The resulting numerical levels were then extrapolated in the truncation energy using the extrapolation scheme of Ref.~\cite{BajnokMarton2}. 

We find that, even after extrapolation, the free-boson TSM does not yield a satisfactory collapse of the extracted two-particle Bethe--Yang lines onto a single phase-shift curve when the bulk energy density and one-particle mass are taken directly from the same free-boson TSM data. A consistent collapse is instead obtained if, for the free-boson TSM, one uses Borel-resummed estimates for the bulk energy and mass. In order to perform an \emph{apples-to-apples} comparison with the variational/integrable-basis (FFTSM) results, we applied the same extrapolation ansatz (i.e., the same fitting form in the truncation energy) to the FFTSM spectra and, when comparing extrapolated data between the two methods, used the same Borel-resummed bulk energy and mass as in the extrapolated free-boson TSM analysis. With this choice, the extrapolated free-boson TSM and extrapolated FFTSM phase-shift extractions are mutually consistent and overlap with the FFTSM curve obtained already \emph{without} extrapolation. Importantly, for FFTSM we also find that using the bulk energy density and mass extracted directly from the FFTSM spectra leads to robust phase-shift lines and a stable collapse, highlighting the improved control over finite-size and truncation effects afforded by the integrable basis.

This comparison shows that the use of an integrable basis for the TSM is not only advantageous in reducing the number of basis states, compared to the non-interacting one, but also in producing accurate estimates of purely interacting quantities, like the scattering phase shifts, without resorting to extrapolations.


The numerical determination of the $S$-matrix phase below threshold using the Truncated Space Method thus constitutes the central non-perturbative result of this work. The corresponding variational parameters $b_g$ and effective $S$-matrix couplings $b'(g)$ for all $g \leq 1/3$ are summarized in Table~\ref{t_couplings}.

\begin{table}[]
\begin{subtable}{0.45\linewidth}
\begin{tabular}{|c|c|c|}
    \hline
    $g$ & $b_g$ & $b'_g$\\
\hline
    0.005 & 0.358182 & 0.3171$\pm$ 0.0033 \\
 0.01 & 0.489686 & 0.4845$\pm$ 0.0027 \\
 0.015 & 0.600453 & 0.6050$\pm$ 0.0025 \\
 0.02 & 0.696739 & 0.7043$\pm$ 0.0023 \\
 0.025 & 0.777461 & 0.7906$\pm$ 0.0022 \\
 0.03 & 0.849712 & 0.8680$\pm$ 0.0021 \\
 0.035 & 0.916382 & 0.9388$\pm$ 0.0021 \\
 0.04 & 0.978593 & 1.0045$\pm$ 0.0020 \\
 0.045 & 1.03701 & 1.0661$\pm$ 0.0020 \\
 0.05 & 1.09219 & 1.1243$\pm$ 0.0020 \\
 0.055 & 1.14457 & 1.1797$\pm$ 0.0020 \\
 0.06 & 1.19448 & 1.2328$\pm$ 0.0021 \\
 0.065 & 1.24218 & 1.2838$\pm$ 0.0021 \\
 0.07 & 1.28792 & 1.3319$\pm$ 0.0012 \\
 0.075 & 1.33187 & 1.3798$\pm$ 0.0012 \\
 0.08 & 1.37418 & 1.4263$\pm$ 0.0013 \\
 0.085 & 1.415 & 1.4717$\pm$ 0.0014 \\
 0.09 & 1.45444 & 1.5162$\pm$ 0.0014 \\
 0.095 & 1.49259 & 1.5597$\pm$ 0.0015 \\
 0.1 & 1.52955 & 1.6026$\pm$ 0.0015 \\
 0.105 & 1.56539 & 1.6448$\pm$ 0.0016 \\
 0.11 & 1.60017 & 1.6864$\pm$ 0.0017 \\
 0.115 & 1.63396 & 1.7276$\pm$ 0.0017 \\
 0.12 & 1.66681 & 1.7684$\pm$ 0.0018 \\
 0.125 & 1.69877 & 1.8088$\pm$ 0.0018 \\
 0.13 & 1.72989 & 1.8490$\pm$ 0.0019 \\
 0.135 & 1.7602 & 1.8890$\pm$ 0.0019 \\
 0.14 & 1.78973 & 1.9288$\pm$ 0.0019 \\
 0.145 & 1.81853 & 1.9686$\pm$ 0.0020 \\
 0.15 & 1.84663 & 2.0082$\pm$ 0.0020 \\
 0.155 & 1.87405 & 2.0479$\pm$ 0.0020 \\
 0.16 & 1.90082 & 2.0876$\pm$ 0.0021 \\
 0.165 & 1.92696 & 2.1273$\pm$ 0.0021 \\
    \hline
\end{tabular}
\end{subtable}
\begin{subtable}{0.45\linewidth}
\begin{tabular}{|c|c|c|}
    \hline
    $g$ & $b_g$ & $b'_g$\\
    \hline
 0.17 & 1.9525 & 2.1672$\pm$ 0.0021 \\
 0.175 & 1.97746 & 2.2072$\pm$ 0.0022 \\
 0.18 & 2.00186 & 2.2474$\pm$ 0.0022 \\
 0.185 & 2.02571 & 2.2879$\pm$ 0.0023 \\
 0.19 & 2.04904 & 2.3286$\pm$ 0.0024 \\
 0.195 & 2.07186 & 2.3695$\pm$ 0.0025 \\
 0.2 & 2.09419 & 2.4108$\pm$ 0.0027 \\
 0.205 & 2.11605 & 2.4524$\pm$ 0.0029 \\
 0.21 & 2.13744 & 2.4944$\pm$ 0.0031 \\
 0.215 & 2.15839 & 2.5369$\pm$ 0.0034 \\
 0.22 & 2.1789 & 2.580$\pm$ 0.004 \\
 0.225 & 2.19898 & 2.623$\pm$ 0.004 \\
 0.23 & 2.21866 & 2.667$\pm$ 0.005 \\
 0.235 & 2.23794 & 2.712$\pm$ 0.005 \\
 0.24 & 2.25683 & 2.757$\pm$ 0.006 \\
 0.245 & 2.27534 & 2.803$\pm$ 0.006 \\
 0.25 & 2.29348 & 2.850$\pm$ 0.007 \\
 0.255 & 2.31127 & 2.897$\pm$ 0.008 \\
 0.26 & 2.32871 & 2.945$\pm$ 0.008 \\
 0.265 & 2.34581 & 2.994$\pm$ 0.009 \\
 0.27 & 2.36257 & 3.044$\pm$ 0.010 \\
 0.275 & 2.37902 & 3.096$\pm$ 0.011 \\
 0.28 & 2.39515 & 3.148$\pm$ 0.012 \\
 0.285 & 2.41097 & 3.201$\pm$ 0.013 \\
 0.29 & 2.4265 & 3.256$\pm$ 0.015 \\
 0.295 & 2.44174 & 3.313$\pm$ 0.016 \\
 0.3 & 2.45669 & 3.371$\pm$ 0.018 \\
 0.305 & 2.47136 & 3.430$\pm$ 0.019 \\
 0.31 & 2.48576 & 3.492$\pm$ 0.022 \\
 0.315 & 2.4999 & 3.556$\pm$ 0.024 \\
 0.32 & 2.51378 & 3.623$\pm$ 0.027 \\
 0.325 & 2.5274 & 3.692$\pm$ 0.030 \\
 0.33 & 2.54079 & 3.766$\pm$ 0.033 \\
    \hline
\end{tabular}
\end{subtable}
    \caption{Summary table of the optimal coupling $b_g$ obtained with the variational method and the $S$-matrix best-fit parameter $b'_g$ obtained through the TSM.}
    \label{t_couplings}
\end{table}

\section{Conclusions}\label{s_conclusions}

In this work, we have established a variational framework that bridges integrable and non-integrable quantum field theories in a controlled and quantitative manner. By employing the sinh–Gordon model as a variational reference for the $\varphi^4$ Landau–Ginzburg theory, we demonstrated that exact knowledge of vacuum expectation values and form factors in an integrable theory can be leveraged to extract non-perturbative information about its non-integrable counterpart.
Our analysis successfully reproduced, with remarkable precision, the vacuum energy density, physical mass, and low-energy scattering properties of the $\varphi^4$ model in the weak-coupling regime. The agreement between the variational estimates, the Borel-resummed perturbative series, and the finite-volume TSM spectra illustrates the power and flexibility of this hybrid variational–numerical approach. In particular, the use of the sinh–Gordon basis within the Truncated Space Method significantly mitigates finite-size artifacts, revealing a deep structural continuity between the two models.

Perhaps most importantly, this study underscores how the crystalline structure of integrable quantum field theories can serve as a guiding principle for navigating the mostly unknown landscape of non-integrable dynamics. The correspondence between the variationally optimized coupling $b_g$ and the effective scattering coupling extracted from finite-volume spectra further demonstrates the internal consistency and physical insight offered by this approach.

An especially intriguing open problem is to extend this framework to the critical region of the $\varphi^4$ theory, where the renormalized mass vanishes and universal scaling sets in. Understanding how the variational principle reorganizes near criticality—whether through the emergence of new effective degrees of freedom or via a breakdown of the sinh–Gordon correspondence—would provide a profound test of the method’s reach and a deeper window into the non-perturbative structure of two-dimensional quantum field theories. Equally interesting would be to extend the variational method to the ${\mathbb Z}_2$ broken phase of the $\varphi^4$ theory. 

We believe that the synthesis of variational, integrable, and numerical methods presented here may open new avenues for addressing non-integrable models, paving the way toward a unified description of quantum field dynamics beyond perturbation theory.

\begin{acknowledgments}
GM and AH acknowledge the grants PNRR MUR Project PE0000023- NQSTI and PRO3 Quantum Pathfinder.
AH is thankful to Pavlo Gavrylenko for the numerous discussions. AH appreciates the suggestion of Zolt\'{a}n Bajnok, Marton Kormos and G\'{a}bor Tak\'{a}cs to apply TSM. 
AS would like to thank Robert Konik for discussions and the BIRS center for their kind hospitality during the workshop ``Exact Solutions in Quantum Information: Entanglement, Topology, and Quantum Circuits'', where parts of this work were completed.
\end{acknowledgments}

\appendix
\section{Setup of the TSM Hamiltonian}\label{a_TSM_explanation}

Building up the truncated basis proceeds as follows. We use a convention
for the phase shift in which it asymptotes to zero at both infinities
of the rapidity. Therefore the $\delta\left(\theta\right)$ is antisymmetric,
with a discontinuity at $\theta=0$. With this choice, the Bethe quantum
numbers are ``bosonic'' (despite the physical particles themselves
being fermionic due to $S(0)=-1$), with repetitions allowed. We first
generate a set of states with right-moving particles only in an essentially
chiral construction. For each integer $0<K\leq K_{max}$, we generate
all integer partitions of $K$, and the numbers appearing in each
decomposition are associated with the Bethe quantum numbers of right-moving
particles. Depending on the momentum sector $P$, these quasi-chiral
momentum sets are sewn together with their negative momentum counterparts.
Finally, new states with extra zero momentum particles are added until
all states below a particle number threshold $N_{max}$ are accounted
for. Finally, we select states with definite $Z_{2}$ parity and auxiliary
energy $E^{\left(0\right)}\leq E_{\text{threshold}}^{\left(0\right)}$.

Additional consideration is given to odd $Z_{2}$ parity states in
the $P=0$ sector with an odd number of zero modes that are invariant
with respect to spatial inversion. These states have a particle with
exactly zero rapidity due to symmetry. This leads to extra disconnected
pieces in the matrix elements between such states, necessitating their
separate treatment.

Once the Bethe quantum numbers are collected, the Asymptotic Bethe
Ansatz is solved to obtain the properly quantized particle rapidities.
The clustering property of form factors is used to simplify the case
of diagonal matrix elements and exactly coinciding rapidities. To
this end, we add an extra particle of large rapidity to one vector
in the matrix element. The extra particle acts as a regulator. It
can be shown that with the appropriate normalization, this method
yields the diagonal matrix elements in the infinite rapidity limit.

Once the rapidities are known, we employ the phase-enhanced version
of the Pozsgay-Tak\'{a}cs finite volume form factor formula  \cite{POZSGAY2008167,POZSGAY2008209,Bajnok:2019cdf}. Application
of the formula requires calculating the Gaudin determinants (see below) corresponding
to each basis state, as well as the appropriate phase factors. Note that the Pozsgay-Tak\'{a}cs formulae provide the finite volume form factors up to (L\"{u}scher) corrections exponentially small at large volume. We restrict to a volume range where these corrections are safely neglected.

To obtain the matrix elements of the perturbation, we use
\begin{equation}
\left\langle \left\{ I_{i}\right\} _{i=1}^{k}\left|\intop_{0}^{R}dx\:O\left(0,x\right)\right|\left\{ \tilde{I}_{j}\right\} _{j=1}^{l}\right\rangle _{R}=R\left\langle \left\{ I_{i}\right\} _{i=1}^{k}\left|O\left(0,0\right)\right|\left\{ \tilde{I}_{j}\right\} _{j=1}^{l}\right\rangle _{R}.
\end{equation}
According to the Pozsgay-Takács formula, for nondiagonal form factors,
inside matrix elements, we are entitled to make the substitution
\begin{equation}
\left|\left\{ \tilde{I}_{j}\right\} _{j=1}^{l}\right\rangle _{R}\rightarrow\tilde{N}_{l}\left(\left\{ \vartheta\right\} \right)\left|\left\{ \vartheta_{j}\right\}_{j=1}^{l}\right\rangle ,
\end{equation}
where the $N_{l}$ is a complex normalization factor
\begin{equation}
\tilde{N}_{l}\left(\left\{ \vartheta\right\} \right)=\left(\rho_{l}\left(\left\{ \vartheta\right\} \right)\mathcal{S}_{l}\left(\left\{ \vartheta\right\} \right)\right)^{-\frac{1}{2}},
\end{equation}
with $\rho_{l}$ is the Gaudin determinant
\begin{equation}
\rho_{l}\left(\left\{ \vartheta\right\} \right)=\det G_{pq},\quad G_{pq}=\partial_{\theta_{p}}Q_{q}\left(\left\{ \vartheta\right\} \right),
\end{equation}
where
\begin{equation}
Q_{q}\left(\left\{ \vartheta\right\} \right)=p_q R+\sum_{\substack{j=1\\ j\neq q}}^n \delta\!\left(\vartheta_q-\vartheta_j\right),\quad q\in\{1,\dots,n\} 
\end{equation}
is the LHS of the asymptotic Bethe Ansatz with $\delta(\theta)=-i\log S(\theta)$,
\begin{equation}
\mathcal{S}_{l}\left(\left\{ \vartheta\right\} \right)=\prod_{s=1}^{k}\prod_{r=1}^{s-1}S\left(\vartheta_{r}-\vartheta_{s}\right).
\end{equation}
In turn, the infinite volume matrix elements can be expressed in terms
of the elementary form factors
\begin{equation}
\left\langle \left\{ \theta_{i}\right\} _{i=1}^{k}\left|O\left(0,0\right)\right|\left\{ \vartheta_{j}\right\} _{j=1}^{l}\right\rangle =F^{O}\left(\left\{ i\pi+\theta_{i}\right\} _{i=1}^{k},\left\{ \vartheta_{j}\right\} _{j=1}^{l}\right).
\end{equation}
Here, the sets of rapidities of the left and right states are denoted by $\{\theta\}$ and $\{\vartheta\}$, respectively.

The elementary form factors of the exponential field in the sinh-Gordon
model are well-known and reproduced in Eq. \eqref{elementary_FF_formula}.

We can write this as
\begin{equation}
\begin{split}
&F^{\exp b\varphi}\left(\left\{ i\pi+\theta_{i}\right\} _{i=1}^{k},\left\{ \vartheta_{j}\right\} _{j=1}^{l}\right)\\
&=\prod_{i<p}^{k}\frac{F_{\text{min}}\left(\theta_{i}-\theta_{p}\right)}{-e^{\theta_{i}}-e^{\theta_{p}}}\prod_{j<q}^{l}\frac{F_{\text{min}}\left(\vartheta_{j}-\vartheta_{q}\right)}{e^{\vartheta_{j}}+e^{\vartheta_{q}}}F_{\text{ren}}^{\exp b\varphi}\left(\left\{ i\pi+\theta_{i}\right\} _{i=1}^{k},\left\{ \vartheta_{j}\right\} _{j=1}^{l}\right),
\end{split}
\end{equation}
\begin{equation}
\begin{split}
&F_{\text{ren}}^{\exp b\varphi}\left(\left\{ i\pi+\theta_{i}\right\} _{i=1}^{k},\left\{ \vartheta_{j}\right\} _{j=1}^{l}\right)\\
&=\langle e^{\beta\varphi}\rangle H_{k+l}\:Q_{k+l}\left(-e^{\theta_{1}},\dots-e^{\theta_{k}},e^{\vartheta_{1}},\dots e^{\vartheta_{l}}\right)\prod_{i=1}^{k}\prod_{j=1}^{l}\frac{F_{\text{min}}\left(\theta_{i}-\vartheta_{j}\right)}{-e^{\theta_{i}}+e^{\vartheta_{j}}},
\end{split}
\end{equation}
where $H_{k+l}=\left(\frac{4\sin(\pi \alpha)}{F_\text{min}(i\pi)}\right)^{\frac{n}{2}}$ is a normalization and $Q_{k+l}=\det M(b)$ is a symmetric polynomial, see \eqref{Q-det} and \eqref{sigma-generator}.
We can finally express the finite volume matrix elements in a numerically
feasible way as
\begin{equation}
\left\langle \left\{ I_{i}\right\} _{i=1}^{k}\left|e^{b\varphi}\left(0,0\right)\right|\left\{ \tilde{I}_{j}\right\} _{j=1}^{l}\right\rangle _{L}=-N_{k}^{*}\left(\left\{ \theta\right\} \right)N_{l}\left(\left\{ \vartheta\right\} \right)F_{\text{ren}}^{\exp b\varphi}\left(\left\{ i\pi+\theta_{i}\right\} _{i=1}^{k},\left\{ \vartheta_{j}\right\} _{j=1}^{l}\right)
\end{equation}
with
\begin{equation}
N_{l}\left(\left\{ \vartheta\right\} \right)=\tilde{N}_{l}\left(\left\{ \vartheta\right\} \right)\prod_{j<q}^{l}\frac{F_{\text{min}}\left(\vartheta_{j}-\vartheta_{q}\right)}{e^{\vartheta_{j}}+e^{\vartheta_{q}}}.
\end{equation}

In the numerics, we use the closed-form expression for $F_{\text{min}}$ \cite{CASTROALVAREDO2024116459} (an alternative form is given by \eqref{fmin}),

\begin{equation}
\begin{aligned}
\log F_{\text{min}}\left(t\right) & =\frac{1}{2}\log2+\log\left(-i\sinh\frac{t}{2}\right)-\frac{1}{4}\log\left[\cosh^{2}\left(t\right)-\sin^{2}\left(\frac{\pi r}{2}\right)\right]\\
 & -\frac{b}{4}\log\left[\frac{\cosh t-\sin\frac{\pi r}{2}}{\cosh t+\sin\frac{\pi r}{2}}\right]+\frac{i\left(t-i\pi\right)}{2\pi}\log\left[\frac{i\cos\frac{\pi r}{2}-\sinh t}{i\cos\frac{\pi r}{2}+\sinh t}\right]\\
 & +\frac{i}{4\pi}\left[\mathrm{Li}_{2}\left(-ie^{t-\frac{i\pi r}{2}}\right)-\mathrm{Li}_{2}\left(ie^{t-\frac{i\pi r}{2}}\right)+\mathrm{Li}_{2}\left(-ie^{t+\frac{i\pi r}{2}}\right)-\mathrm{Li}_{2}\left(ie^{t+\frac{i\pi r}{2}}\right)\right.\\
 & \left.+\mathrm{Li}_{2}\left(-ie^{-t-\frac{i\pi r}{2}}\right)-\mathrm{Li}_{2}\left(ie^{-t-\frac{i\pi r}{2}}\right)+\mathrm{Li}_{2}\left(-ie^{-t+\frac{i\pi r}{2}}\right)-\mathrm{Li}_{2}\left(ie^{-t+\frac{i\pi r}{2}}\right)\right],
\end{aligned}
\end{equation}
where $\operatorname{Li}$ is the dilogarithm function, $r=1-2\alpha$ and $\alpha$ is defined in      Eq.~\eqref{eq_shG_smatrix}.

Finally, we obtain the matrix elements of the normal ordered powers of the field $\varphi$ by differentiating the vertex operators. The relevant formulae are collected below. We first introduce a relative form factor as building block,

\begin{equation}
\mathrm{FF}^{:\varphi^n:}_\text{rel}=H_{k+\ell}\:\prod_{i=1}^k\prod_{j=1}^{\ell} \frac{ F_{min}\left(\theta_i-\vartheta_j\right)}{x_{i}+x_{j}}\Phi\left(x,n\right),
\end{equation}
where
\begin{equation}
\begin{aligned}
\Phi\left(x,1\right) & =\mathbb{D}\:\mathrm{tr}V_{1},\\
\Phi\left(x,2\right) & =\mathbb{D}\left[\left(\mathrm{tr}V_{1}\right)^{2}+\mathrm{tr}W_{2}\right],\\
\Phi\left(x,3\right) & =\mathbb{D}\left[\left(\mathrm{tr}V_{1}\right)^{3}+3\mathrm{tr}V_{1}\mathrm{tr}W_{2}+\mathrm{tr}W_{3}\right],\\
\Phi\left(x,4\right) & =\mathbb{D}\left[\left(\mathrm{tr}V_{1}\right)^{4}+6\left(\mathrm{tr}V_{1}\right)^{2}\mathrm{tr}W_{2}+3\left(\mathrm{tr}W_{2}\right)^{2}+4\mathrm{tr}V_{1}\mathrm{tr}W_{3}+\mathrm{tr}W_{4}\right].
\end{aligned}
\end{equation}
Introducing the notation
\begin{equation}
d(k)_{ij}=\left(-1\right)^{2i-j}\sigma\left(2i-j\right)\left[i-j+\frac{a}{b}\right]^{\left(k\right)}
\end{equation}
(with $i,j=1,\dots$) where
\begin{equation}
\left[n\right]^{\left(k\right)}=\left(\frac{\alpha\pi}{b}\right)^{k}\frac{\sin\left(n\pi\alpha+\frac{k\pi}{2}\right)}{\sin\left(\pi\alpha\right)},
\end{equation}
the variable $\mathbb{D}$ is defined as
\begin{equation}
\mathbb{D}=\begin{cases}
\mathrm{det}\:d(0), & \mathrm{dim}\:d(0)>1,\\
1, & \mathrm{dim}\:d(0)=1.
\end{cases}
\end{equation}

Furthermore, the quantities $W_i$ are defined as
\begin{align}
W_{2} & =-V_{1}^{2}+V_{2},\\
W_{3} & =-2V_{1}W_{2}-V_{2}V_{1}+V_{3},\\
W_{4} & =-3V_{1}W_{3}-3V_{2}W_{2}-V_{3}V_{1}+V_{4},
\end{align}
and $V_n$ takes the form
\begin{align*}
V_{n} & =d(0)^{-1}d(n).
\end{align*}

Since the VEV of $e^{a\varphi}$ also depends on the variable $a$, extracting powers of the field by differentiating the vertex operators requires applying the product rule. We write the renormalized vertex-operator form factors schematically as
\begin{equation}
F^{e^{a\varphi}}_{\rm ren}(\{\theta\},\{\vartheta\})
=\langle e^{a\varphi}\rangle\;\mathrm{FF}^{e^{a\varphi}}_{\rm rel}(\{\theta\},\{\vartheta\}),
\end{equation}
where $\mathrm{FF}_{\rm rel}$ contains the normalization $H_{k+\ell}$, the minimal form factors and the symmetric-polynomial part. Differentiating at $a=0$ then produces mixing with the identity through the derivatives of the VEV. In particular, introducing
\begin{equation}
\mathcal{H}(a)=\langle e^{a\varphi}\rangle\,[a/b],
\end{equation}
so that $\mathcal{H}(a)$ collects the $a$-dependence of the overall prefactor, we obtain for the (non-normal-ordered) even powers
\begin{equation}
\label{phi-2-updated}
F_{\rm ren}^{\varphi^2}(\{\theta\},\{\vartheta\})
=
\mathrm{FF}^{:\varphi^2:}_{\rm rel}(\{\theta\},\{\vartheta\})
+\partial_a^2\mathcal{H}(0)\:\mathbf{1},
\end{equation}
\begin{equation}
\label{phi-4-updated}
F_{\rm ren}^{\varphi^4}(\{\theta\},\{\vartheta\})
=
\mathrm{FF}^{:\varphi^4:}_{\rm rel}(\{\theta\},\{\vartheta\})
+6\,\partial_a^2\mathcal{H}(0)\:\mathrm{FF}^{:\varphi^2:}_{\rm rel}(\{\theta\},\{\vartheta\})
+\partial_a^4\mathcal{H}(0)\:\mathbf{1},
\end{equation}
where $\mathbf{1}$ denotes the identity operator contribution (equivalently the finite-volume Kronecker delta in the truncated basis), and the coefficient $6$ is the combinatorial factor arising from the product rule.

The formulae above apply directly to non-diagonal matrix elements, i.e.\ when the sets
$\{\theta\}$ and $\{\vartheta\}$ have no exactly coinciding rapidities after crossing.
In the diagonal case, and more generally in the presence of exact coincidences (including
the situation discussed in Ref.~\cite{POZSGAY2008209}), additional disconnected pieces
appear. Rather than implementing the full disconnected-term subtraction explicitly, in the
numerics we employ a regulator based on the clustering (asymptotic factorization) property
of sinh-Gordon form factors: we add to one side of the matrix element an auxiliary particle
with very large rapidity $\Theta\gg 1$ (equivalently a very large Bethe quantum number),
solve the corresponding Bethe--Yang equations, and evaluate the resulting \emph{non-diagonal}
finite-volume form factor using the Pozsgay--Tak\'{a}cs prescription. In the limit
$\Theta\to +\infty$ the extra particle factorizes and one recovers the desired diagonal /
coincident-rapidity matrix element (up to exponentially small L\"uscher corrections), with
the auxiliary particle acting as a regulator that detunes the would-be kinematical poles.
We note that this procedure changes the particle-number parity on the regulated side,
which must be accounted for when selecting the appropriate $Z_2$ sector in the numerical
implementation.

In the actual numerical implementation we further exploit that the determinant-based object
\texttt{getmat} returns simultaneously the derivatives of the symmetric polynomial $Q_{k+\ell}$ with respect
to the vertex-operator parameter (encoded through the $k$-index in $d(k)$ and implemented via \texttt{sbraDHi}).

The vector returned by our determinant construction (implemented by \texttt{FFphi}/\texttt{getmat}) should be viewed as encoding charge-derivatives of the \emph{determinant/polynomial block} $G(a)$ of the vertex-operator form factor, rather than the Taylor coefficients of the full operator $e^{a\varphi}$ by itself. With our conventions the full charge-dependent prefactor factorizes as
\begin{equation}
F^{e^{a\varphi}}_{\rm ren}(\{\theta\},\{\vartheta\})
=\mathcal H(a)\,G(a),\qquad 
\mathcal H(a)=\langle e^{a\varphi}\rangle\,[a/b],
\end{equation}
where $\langle e^{a\varphi}\rangle$ is an even function of $a$ while the sine-ratio block $[a/b]$ is odd, hence $\mathcal H(a)$ is odd and in particular $\mathcal H(0)=0$. As a consequence, the product rule implies that even derivatives of the full form factor at $a=0$ involve odd derivatives of $G$:
\begin{equation}
\frac{d^{2}}{da^{2}}F^{e^{a\varphi}}_{\rm ren}\big|_{a=0}
=2\,\mathcal H'(0)\,G'(0),\qquad
\frac{d^{4}}{da^{4}}F^{e^{a\varphi}}_{\rm ren}\big|_{a=0}
=4\,\mathcal H'(0)\,G^{(3)}(0)+4\,\mathcal H^{(3)}(0)\,G'(0).
\end{equation}
This is the origin of the reconstruction used in the numerics: for generic (non-coincident) matrix elements the operators $\varphi^{2}$ and $\varphi^{4}$ are assembled from the \emph{odd} derivative components of the \texttt{FFphi} vector (corresponding to $G'(0)$ and $G^{(3)}(0)$), with known prefactors determined by $\mathcal H'(0)$ and $\mathcal H^{(3)}(0)$.  When the auxiliary-particle
regulator is used, particle-number parity on the regulated side is flipped, and the mapping shifts by one:
the same operators are then assembled from the \emph{even} derivative components (those labeled $0,2,4$),
together with the VEV insertions according to Eqs.~\eqref{phi-2-updated}--\eqref{phi-4-updated}. 

Finally, in
practice we evaluate these derivatives at a small but nonzero parameter $a=\texttt{apar}\ll 1$ (rather than
exactly at $a=0$) as a numerical regulator; convergence in $a$ can be checked by varying \texttt{apar} and the
working precision.

\clearpage
\bibliographystyle{elsarticle-num}
\bibliography{paper.bib}

\end{document}